\documentclass[authoryear]{elsarticle}

\usepackage{geometry}
\usepackage[hidelinks]{hyperref}

\usepackage{nccmath}

\usepackage{amsfonts} 
\usepackage{relsize}

\usepackage{algorithm}
\usepackage{mathtools}
\usepackage{amsmath}
\usepackage[noend]{algpseudocode}
\usepackage{array}
\usepackage{booktabs}

\usepackage{graphicx}
\graphicspath{{Images/}}
\usepackage{subcaption}
\usepackage[table,xcdraw]{xcolor}

\usepackage{tabularx}
\newcolumntype{Y}{>{\centering\arraybackslash}X}

\journal{arXiv}

\begin{document}

\begin{abstract}
Linguistic labels are effective means of expressing qualitative assessments because they account for the uncertain nature of human preferences. However, to perform computations with linguistic labels, they must first be converted to numbers using a scale function. Within the context of the Analytic Hierarchy Process (AHP), the most popular scale used to represent linguistic labels numerically is the linear 1-9 scale, which was proposed by Saaty. However, this scale has been criticized by several researchers, and various alternatives are proposed in the literature. There is a growing interest in scale \textit{individualization} rather than relying on a generic fixed scale since the perceptions of the decision maker regarding these linguistic labels are highly subjective. The methods proposed in the literature for scale individualization focus on minimizing the transitivity errors, i.e., consistency. In this research, we proposed a novel, easy-to-learn, easy-to-implement, and computationally less demanding scale individualization approach based on compatibility. We also developed an experimental setup and introduced two new metrics that can be used by researchers that contribute to the theory of AHP. To assess the value of scale individualization in general, and the performance of the proposed novel approach in particular,  numerical and two empirical studies are conducted. The results of the analyses demonstrate that the scale individualization outperforms the conventional fixed scale approach and validates the benefit of the proposed novel heuristic.

\label{Abstract}
\end{abstract}

\begin{keyword}
\texttt{Decision Processes\sep Linguistic Labels \sep Scale Individualization \sep Pairwise Comparisons \sep Multi Criteria Decision Making (MCDM)} 
\end{keyword}

\begin{frontmatter}

\title{A Basic Algorithm for Generating Individualized Numerical Scale (BAGINS)}

\author{Faran Ahmed\fnref{Faran}}
\author{Kemal Kilic\fnref{Kemal}}
\address{Sabanci University, Orta Mahalle, Universite Caddesi, No:27, 34956 Tuzla, Istanbul \\ National University of Science and Technology (NUST), 44000, Islamabad, Pakistan}
\fntext[Faran]{Assistant Professor, NUST Business School, faran.ahmed@nbs.nust.edu.pk}
\fntext[Kemal]{Associate Professor, Sabanci University, kkilic@sabanciuniv.edu}

\end{frontmatter}

\section{Introduction}
\label{Introduction}
Humans prefer using linguistic labels as opposed to numbers in order to express their opinions since uncertainty associated with their preferences is better communicated using vague verbal terms that are more intuitive and natural \citep{mandel2021arithmetic}. Due to the uncertain nature of the entire decision problem, precise numerical values are generally avoided because they may imply a sense of precision that a decision-maker does not want.  \citep{huizingh1997comparison}. For example, people mostly think and talk about uncertainty in terms of verbal phrases (e.g., \textit{likely}, \textit{almost certainly not}, etc.,) and are more skilled in using the rules of language as compared to employing the rules of probability \citep{budescu1985consistency}. 

Linguistic labels are used across different domains such as business, academia, intelligence, medicine and politics in order to express preferences and/or judgments (e.g., significance, probabilities, etc.). Some examples of linguistic labels are tabulated in Table~\ref{Lexicons}. The perceptions of the decision-makers regarding these linguistic labels are highly subjective, i.e., they have different meanings for different individuals. For example, the pioneering work of Sherman Kent \citep{kent1964words} regarding the perception of 11 probability phrases demonstrates that when the intelligence officers are asked to quantify these phrases, there is a significant degree of deviation associated with each one of the phrases (e.g., \textit{Probable} corresponds to $75 \% \pm12 \%$, \textit{Almost Certain} corresponds to $93 \% \pm 6 \%$, etc.). A similar study which is conducted with thirty financial strategy experts also concludes that the perceptions of probability phrases are subjective and vary across different individuals \citep{tavana1997applied}.

\begin{table}[ht!]
\small
\centering
\caption{Linguistic labels across different domains} 
\vspace*{-3mm}
\label{Lexicons}
\begin{tabular}{ll}   
\hline
Analytic Hierarchy Process (AHP) \citep{saaty1977scaling} & Numerical Equivalent \\    
\hline
Equal Importance & 1 \\
Moderate Importance & 3  \\ 
Strong Importance & 5 \\ 
Very Strong Importance & 7 \\
Extreme Importance & 9\\
\hline
Words of Estimate Probability in Intelligence \citep{kent1964words} & Percentage Equivalent \\
\hline
Almost Certain & 93\% $(\pm 6 \%)$\\
Probable & 75\% $(\pm 12 \%)$ \\
Chance About Even & 50\% $(\pm 10 \%)$\\
Probably Not & 30\%  $(\pm 10 \%)$\\
Almost Certainly Not & 7\% $(\pm 5 \%)$\\
\hline
Words of Estimate Probability in Medicine \citep{WEPmedecine} & Quantitative Equivalent \\
\hline
Likely & Expected to happen to more than 50\% of subjects \\
Frequent & Will probably happen to 10-50\% of subjects \\
Occasional & Will happen to 1-10\% of subjects\\
Rare & Will happen to less than 1\% of subjects \\
\hline
\end{tabular}
\end{table}

In this research, we focus on the set of linguistic labels used in the realm of Multi-Criteria Decision Making (MCDM), in particular used as part of the Analytic Hierarchy Process (AHP) proposed by Thomas L. \cite{saaty1977scaling}. Being one of the most popular methods in MCDM, this technique has found diverse applications such as the selection of cloud computing service provider \citep{tanoumand2017selecting}, formulating cryptocurrency mining strategies \citep{hacioglu2021crafting}, decision support system for real-time ambulance relocation \citep{hajiali2022interactive}, selection of agricultural irrigation systems \citep{veisi2022application}, task-oriented crowdsourcing recommendations \citep{li2021social}, etc. Detailed and structured reviews of the various developments of AHP are available in  \cite{ishizaka2011review} and \cite{emrouznejad2017state}. 

In AHP, expert preferences are elicited in the form of pairwise comparisons. Pairwise comparison is the preferred way of eliciting human preferences as the process deals with binary evaluations and is an easier cognitive task when compared to simultaneously evaluating all objects \citep{choo2016mathematical}. In pairwise comparisons, two objects are evaluated at a time and preference intensities are provided in the form of linguistic labels. One of the main challenges lies in computing with these linguistic labels. Generally, the linguistic labels are transformed into numbers and the computations are carried out with these numbers. Both in practice and academic research the fixed generic scale proposed by Thomas L. \cite{saaty1977scaling} is the most popular approach that is used for this purpose. Although the linear scale (i.e., Saaty scale of 1-9) is the most popular scale, it has been criticized by a number of researchers and various alternative generic fixed scales are proposed. Some of the popular scales are tabulated in Table \ref{NumericalScales}.     

\begin{table}[ht!]
\small
\centering
\caption{Numerical scales to quantify lexicons} 
\vspace*{-3mm}
\label{NumericalScales}
\begin{tabular}{lll}   
\hline
Scale & Mathematical Formulation & Parameters \\    
\hline
Linear \citep{saaty1977scaling} & $s = x$ & $x = {1,2,...,9}$ \\
Power \citep{harker1987theory} & $s = x^2$ & $x = {1,2,...,9}$ \\ 
Root Square \citep{harker1987theory} & $s = \sqrt{x}$ & $x = {1,2,...,9}$ \\ 
Geometric \citep{lootsma1993scale} & $s = (\sqrt{2})^{x-1}$ & $x = {1,2,...,9}$ \\
Asymptotic \citep{dodd1995comparison} & $s = tanh^{-1}\big( \frac{\sqrt{3}(x-1)}{14}\big)$ & $x = {1,2,...,9}$ \\
Balanced \citep{salo1997measurement} & $s = \frac{x}{1-x}$ & $x = {0.5,0.55,0.6,...,9}$\\
Logarithmic \citep{ishizaka2011influence}& $s = log_{2}(x+1)$ & $x = {1,2,...,9}$\\

\hline
\end{tabular}
\end{table}  

Representing linguistic labels with any of the \textit{fixed} generic scales tabulated in Table \ref{NumericalScales} is also disputable. It can be claimed that the numerical interpretation of linguistic labels \textit{is not same} for all individuals due to the fact that words possess different meanings for different individuals. It is important for practical and theoretical reasons to evaluate this claim in the context of AHP, analogous to the case of probabilistic phrases (e.g., \citep{kent1964words}; \citep{tavana1997applied}; \citep{budescu1985consistency}. Unfortunately, not many empirical studies are available in the literature. Occasionally, researchers refer to \cite{huizingh1997comparison} to substantiate this claim. However, in that study the authors \textit{do not} provide empirical evidence of variation in numerical interpretations of linguistic labels, but rather they explore the \textit{consequence} of different interpretations on the quality of the AHP analysis. 

The actual empirical evidence which supports the claim that the linguistic labels used during the pairwise comparison process have different meanings for different individuals is provided by \cite{poyhonen1997experiment}. They explored the relationship between the linguistic labels and the numbers through an experiment in which 61 participants adjusted the heights of the two bars to represent a certain linguistic label. For example, when shown the label \textit{moderately larger}, participants adjust the height of one bar to the other such that one bar is \textit{moderately larger} than the other bar. The results of the experiment demonstrate that the numerical counterpart for the linguistic label \textit{moderately larger} is $1.30 \pm 0.24$, \textit{strongly larger} is $2.02 \pm 0.52$ and \textit{very strongly larger} is $3.65 \pm 1.89$. Thus concluding that representative numerical counterparts for the verbal expressions vary across individuals (even to a degree where they overlap with each other e.g., someone's \textit{very strongly larger} is smaller than someone else's \textit{strongly larger}).

Once the scale is chosen, the linguistic pairwise comparison matrices are transformed into numerical matrices via the chosen scale in order to carry out the rest of the analysis. Therefore, the chosen scale directly influences the performance of AHP, and selecting the appropriate scale needs to be carried out diligently.  \cite{finan1999transitive} motivated from a compelling experiment asserts that strict reliance on the Saaty scale can induce artificial inconsistency and harm the performance of the analysis. As a result, instead of the Saaty scale, they propose the use of a geometric scale, which can be calibrated with a parameter in order to induce as little inconsistency as possible by preferences solicited with the verbal scale. This parameter is determined for each pairwise comparison matrix separately, with a process that is conducted aside from the pairwise comparisons. In this process, which is referred to as \textit{transitive calibrations}, the decision maker also provides the interpretation of \textit{moderately important} by means of a percentage (denoted as \textit{s}), and the rest of the scale is computed via the geometric progression. The parametric nature of the proposed scale, which can be calibrated for each pairwise comparison matrix, entitles \cite{finan1999transitive} as the first step towards the \textit{individualization} of the scales in AHP to the best of our knowledge.  \cite{liang2008mapping} also developed a parametric scale that spans a continuum of various fixed scales. The parameter of the developed scale is determined for each decision maker based on the user-specified degree of tolerance to inconsistency ($\varepsilon$). According to their approach, given the tolerance parameter and pairwise comparisons, one can obtain transitivity inequalities, and the corresponding parameter for the individualized scale is obtained as a solution to these inequalities.     

In the literature, the pioneering work that directly addresses the individualization of the scale for AHP is \cite{dong2013numerical}. Based on the transitive calibration concept of \cite{finan1999transitive} and 2-tuple linguistic modeling of \cite{herrera20002}, they propose a non-linear programming model that quantifies the linguistic labels at an individual level. For each linguistic label (e.g., for Saaty Scale, there are 17 such linguistic labels, i.e., those that have the values of 1 to 9 and their reciprocals) they form a theoretical transitive calibration matrix ($17\times 17$). Each individual entry of the theoretical matrix is a linguistic label (or \textit{null}) which has a value equal to the transitive calibration, i.e., the multiplication of the values of the linguistic labels that are associated with the corresponding row and the column. Whenever the resulting value of multiplication is beyond the spectrum of the scale, the entry is set to be \textit{null}. In the example provided in the paper, they use the Saaty Scale as the gold standard in the determination of the theoretical transitive calibration matrix. Another $17\times 17$ matrix is constructed from the \textit{elicited} comparison matrices which represent the \textit{individual characteristic}s of the decision maker. In this matrix, the entries are the elicited (i.e., observed) linguistic labels corresponding to the transitive calibrations for each pair of linguistic labels associated with the corresponding row and the column. The overall objective is to determine the individualized scale that minimizes the total deviation between the theoretical and the observed transitive calibration matrices in order to achieve a \textit{consistent} matrix. In the paper, a numerical and/or empirical study is not presented that demonstrates the performance of the proposed approach, but two illustrative examples are provided instead. 

\cite{zhou2018analytic} also links individualization of the numerical scale to the \textit{consistency} of the numerical preferences and proposes a two-stage consistency-driven optimization model to individualize the numerical scale. In the first stage, the pairwise comparison matrices that are formed with linguistic labels are considered and their consistencies are checked based on \textit{transitive properties}. In the second stage, a mathematical programming model that minimizes the inconsistency index of the numerical pairwise comparison matrices is used in order to determine the individualized scale. The mathematical model's objective is to minimize the sum of the normalized log deviations from the transitive calibrations for each observed pairwise comparison. The decision variables, i.e., the individualized scales, are constrained within a particular interval around the corresponding Saaty scale. The reciprocal property and monotonicity among the linguistic labels are also observed as constraints of the model. The model was later transformed into a linear programming model in order to be solved easier. A numerical study is conducted in order to compare the performance of the proposed approach with the Saaty scale. 

With the advance of digital transformation in all aspects of our daily lives, digital assistants such as Apple's Siri, Google's Assistant, Amazon's Alexa, etc. are becoming part of our families. These digital assistants act as predictive chat-bots that use machine learning, natural language processing \& understanding. They learn from the user's preferences, judgments, decisions, and from that learning, they personalize their interactions and provide recommendations to the users. Apparently, in the not-so-distant future, Human-Machine teams would be a more common reality in our business lives as well. Our machine partners should understand what we actually mean when we use a particular linguistic label. There is a need for individualization since linguistic labels have different meanings for different individuals. The individualization of the scales is receiving more attention from the researchers (e.g., \cite{dong2013numerical}, \cite{zhou2018analytic}). However, there is still room for alternative approaches. First of all, the available alternatives are based on mathematical programming and not only the practitioners but also many researchers usually lack the theoretical knowledge to understand and implement them. They are also computationally demanding approaches. On the other hand, they focus on the transitivity errors, i.e., consistency, however other approaches might also be applicable to the individualization process. There is also a need for numerical and empirical studies that assess the value of individualization of the numerical scales in the context of AHP. Note that, unfortunately, there is particularly a lack of empirical studies in the literature. This scarcity has been pointed out by several researchers and such empirical studies are considered as valuable contributions (e.g., \cite{brunelli2018survey}, \cite{cavallo2019comparing},
 \cite{sato2022inconsistency}, ).

In this study, we focus on the individualization of the numerical scales in AHP and address the gaps that are mentioned above. The major contributions of this research are as follows:

\begin{enumerate}
\setlength\itemsep{0em}
    \item A novel heuristic approach, which is easy-to-learn, easy-to-implement, and computationally less demanding that can be used for individualization of the numerical scales in AHP is introduced.
    \item Numerical and empirical studies are conducted to assess the value of individualized scales in the context of AHP and test the performance of the proposed heuristic. 
    \item An experimental analysis framework is developed that can be adopted particularly by the researchers who are contributing to the theory of AHP (e.g., new scales, new priority vector derivation techniques, etc.), and two new performance metrics are proposed.
\end{enumerate}

The remainder of this article is arranged as follows. Section~\ref{Preliminaries} discuss the basic definitions and concepts. Section~\ref{Heuristic} introduces the proposed heuristic, presents an illustrative example regarding its implementation, and provides the motivation behind it. In section~\ref{ResearchMethodology}, we present the research methodology that is adopted to compare the performance of the proposed heuristic with the alternatives. Both the details of the numerical study and the two empirical studies are provided in this section. Results and discussions are presented in section~\ref{ResultsAndDiscussion}. Section~\ref{Conclusions} concludes this research and highlights the future research areas.	

\section{Preliminaries}
\label{Preliminaries}
Let's introduce the basic definitions and knowledge required to understand the rest of the discussions.  

\textbf{Definition 1: Linguistic Pairwise Comparison Matrix (LPCM)} Let $S = (S_k \mid k = 1,2,...,m )$ be a tuple with odd cardinality such that $S_k$ is a linguistic label and $S_i > S_j \text{ if and only if } i > j$ \citep{herrera20002}. The linguistic Pairwise Comparison Matrix LPCM, $L = (l_{ij})_{n\times n}$ represents pairwise comparisons provided by the decision maker, where $l_{ij} \in S$ for $i,j = 1,2,...,n $.

\textbf{Definition 2: Numerical Pairwise Comparison Matrix (NPCM)} Let $A = (a_{ij})_{n \times n}$ be a NPCM such that $a_{ij} >0$ and $a_{ij} \times a_{ji} =1$ \citep{saaty1977scaling} and elements of the matrix $A = (a_{ij})_{n \times n}$ represent preference intensity of the \textit{$i^{th}$} criteria when compared with the \textit{$j^{th}$} criteria. Note that, $A$ is a positive reciprocal matrix.

\textbf{Definition 3: Scale Function} Let $f_{(scale)}$ be the function that maps the linguistic labels $\in S$ to $a_{ij}\in\mathbb{R^{+}}$. Given a numerical scale, function $f_{(scale)}$ transforms linguistic labels into numbers so that $A = f_{(scale)}(L)$. For example, $f_{(Saaty)}$ will transform all linguistic labels into numbers using a scale of 1-9. On the other hand, the inverse scale function $f_{(scale)}^{-1}$ converts all numeric numbers into the corresponding linguistic labels so that $L = f^{-1}_{(scale)}(A)$. Furthermore, $f_{(scale)}^{k}$ denotes the real positive number corresponding to the $k^{th}$ linguistic label (i.e., $S_k$) for a particular scale.  

\cite{vlaev2011does} provides an extensive list of models proposed in the domain of the theory of choice and maps them to three broad categories, i.e., \textit{Value First View}, \textit{Comparison Based Theories} and \textit{Comparison Based Theories without Internal Scale}. According to this classification, well-known decision theories such as Utility Theory \citep{morgenstern1953theory} and Prospect Theory \citep{Kahneman1979Mar} fall under the category of \textit{Value First View}. According to \textit{Value First View}, the brain computes the value of alternatives and favors the ones with higher values. AHP assumes an underlying additive multi-attribute utility function in order to rank the alternatives in a multi-criteria setting. That is to say, according to AHP, a \textit{true priority vector} exists that corresponds to the preferences of the decision maker in a particular context. 

\textbf{Definition 4: True Priority Vector} $V=[v_1, v_2,...,v_n]$ is the true priority vector that corresponds to the preference of the decision maker in a particular context.     

\textbf{Definition 5: Calculated Priority Vector} 
AHP utilizes $L = (l_{ij})_{n\times n}$ and determines the priority vector $w=[w_1, w_2,...,w_n]$ which corresponds to the weight of the criteria or score of the alternatives in terms of each criterion. \cite{saaty1977scaling} proposes that $w$ is the eigenvector associated with the maximum eigenvalue ($\lambda_{max}$) of the  $A = f_{(Saaty)}(L)$ matrix. After the pioneering work of Saaty, some other approaches for priority vector derivation is also proposed in the literature (e.g., Logarithmic Least Squares Method (LLSM) \citep{crawford1985note}, Mean of Normalized Values Heuristics, etc.).  

\begin{figure}[ht!]
\centering
\includegraphics[width=4.1cm, height=5.5cm]{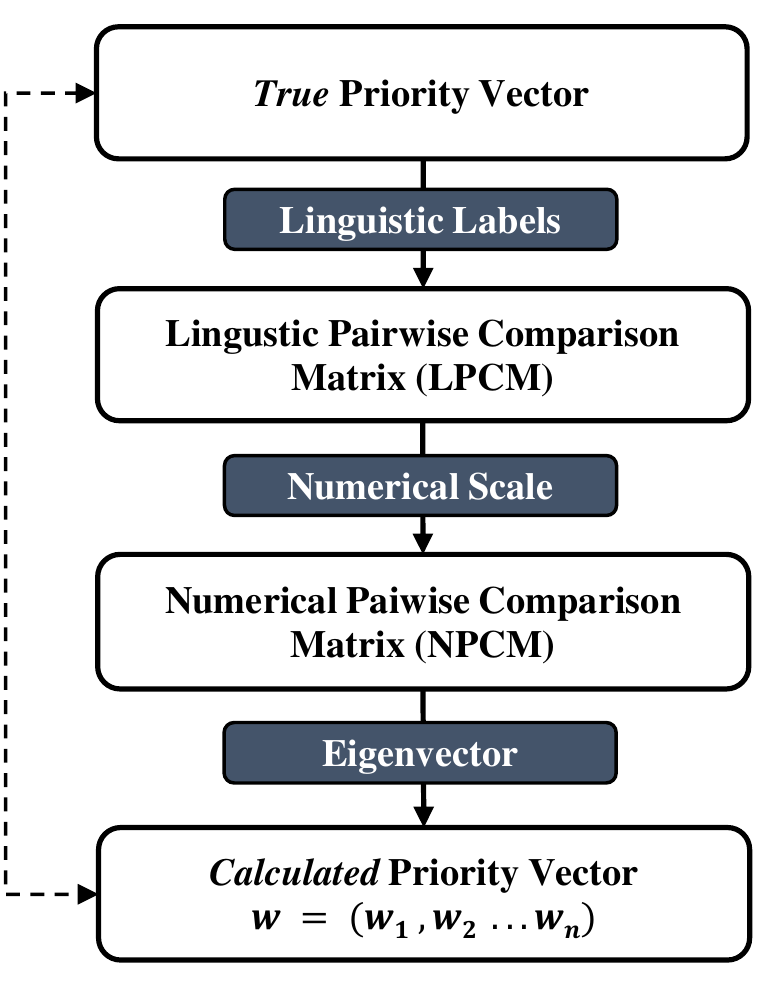}
\caption{Analytic Hierarchy Process}
\label{AHP}
\end{figure}

AHP aims to determine the \textit{true} priority vector via the pairwise comparisons elicited from the decision-makers. This process is illustrated in Figure~\ref{AHP}.   

\textbf{Definition 6: Consistency Index of NPCM} Consistency Index denoted by  $CI$ is defined by Saaty as follows:

\begin{equation}\label{CI_lambdambax}
CI =\frac{\lambda_{max}-n}{n-1} 
\end{equation}

\textbf{Definition 7: Consistency Ratio of NPCM} Let $RI$ be the Random Index which is the average of the $CI$s of the randomly generated NPCMs, then Consistency Ratio is defined by Saaty as $CR = \frac{CI}{RI}$. If $A$ is \textit{sufficiently consistent} then $CR\leq0.1$ and if $A$ is \textit{fully consistent}, then $CR=0$ \citep{saaty1977scaling}. 

In order to address the consistency of NPCM, various other indices are proposed in the literature in addition to the one proposed by Saaty. Comprehensive reviews of various inconsistency indices are available in \cite{brunelli2018survey} and \cite{sato2022inconsistency}.

\textbf{Definition 8: Theoretical Numerical Pairwise Comparison Matrix (TNPCM)} Let priority vector $w=[w_1, w_2,...,w_n]$ be calculated from $A = (a_{ij})_{n \times n}$. Then $W = (w_{ij})_{n \times n}$ is referred to as the theoretical numerical pairwise comparison matrix of $A$. 

\begin{equation}
W = 
\begin{pmatrix}
w_{1}/w_{1} & w_{1}/w_{2} & \cdots & w_{1}/w_{n} \\
w_{2}/w_{1} & w_{2}/w_{2} & \cdots & w_{2}/w_{n} \\
\vdots  & \vdots  & \ddots & \vdots  \\
w_{n}/w_{1} & w_{n}/w_{2} & \cdots & w_{n}/w_{n}
\end{pmatrix}  
\label{Wmatrix}
\end{equation}

\textbf{Definition 9: True Theoretical Numerical Pairwise Comparison Matrix (TTNPCM)} Let priority vector $v=[v_1, v_2,...,v_n]$ be the true priority vector. Then $V = (\frac{v_i}{v_j})_{n \times n}$ is referred to as the true theoretical pairwise comparison matrix. 

\textbf{Definition 10: Compatibility Index Value} The Hadamard product (i.e., element-wise product) of matrix $A$ and $W^{T}$ is represented as $e^{T} A \circ W^{T} e$, where $e$ is the vector of ones of size $n$. Then, the Compatibility Index Value (CIV) is defined as follows \citep{saaty1994ratio};

\begin{equation}\label{CIV}
CIV = n^{-2} . e^{T} A \circ W^{T} e 
\end{equation}

Note that $e^{T}Xe$ for any matrix $X$ would be equal to the sum of all of the elements of the matrix $X$. Thus, we can explicitly write Equation \ref{CIV} as follows: 

\begin{equation}\label{CIV_sum}
CIV = \frac{1}{n^2} \sum_{i=1}^{n} \sum_{j=1}^{n} a_{ij}.\frac{w_j}{w_i}
\end{equation}

A \textit{fully consistent} matrix $A$ and the theoretical pairwise comparison matrix $W$ constructed from the priority vector $w$ of $A$ is exactly same. Therefore, the CIV of $A$ and $W$ would be equal to 1 (see Equation \ref{CIV_sum}). For matrix  $A$ which is inconsistent, CIV will be greater than 1 \citep{saaty1994ratio}.

In the framework of Saaty, where the weights are determined to be the principal right eigenvectors (i.e., $A \times w = \lambda_{max} \times w$), one can easily show that $CIV = \frac{\lambda_{max}}{n}$ (Due to the definition of eigenvector: $\sum_{j=1}^n a_{ij}.w_j=\lambda_{max}.w_i$). Therefore,

\begin{equation}\label{CIV_lambdamax}
CIV=n^{-2} . e^{T} A \circ W^{T} e = \frac{\lambda_{max}}{n}
\end{equation}

Recall that $CI$ deals with a single matrix (e.g., $A$) and measures how much the transitivity of the preferences is violated. On the other hand, CIV deals with the deviation between two matrices (e.g., $A$ and $W$) that represent two priority vectors. However, As Equation \ref{CI_lambdambax} and Equation \ref{CIV_lambdamax} hints, $CI$ and CIV are interrelated measures. The relation between consistency and compatibility is nicely put forward by \cite{saaty1994ratio} as:\textit{``...Consistency is concerned with the compatibility of a matrix of the ratios constructed from a principal right eigenvector with the matrix of judgments from which it is derived.''}. 
This interrelation also can be mathematically observed as follows:

\begin{equation}\label{CIV_CI}
CI = \frac{\lambda_{max}-n}{n-1} \Longrightarrow \lambda_{max} = (n-1).CI+n \Longrightarrow CIV = \frac{(n-1)}{n}.CI+1
\end{equation}

In practice, the true priority vectors are unknown. Therefore, all we can use is the $A$ matrix and/or the associated $W$ matrix. As the above discussion indicates, analysis based on CI or CIV would actually be identical. However, in order to compare the performance of new proposals to the theoretical framework of AHP we can design experimental studies (numerical and/or empirical) in which we have the true priority vectors as well. Therefore, we are going to introduce a \textit{generic} compatibility index which we can use as a performance metric in such experimental analysis.   

\textbf{Definition 11: Generic Compatibility Index Value (GCIV)} Let $X$ and $Y$ be two $n\times n$ matrices. The Hadamard product of matrix $X$ and $Y^{T}$ is represented as $e^{T} X \circ Y^{T} e$. Then, the GCIV is formulated as follows;

\begin{equation}
GCIV = n^{-2} . e^{T} X \circ Y^{T} e 
\end{equation}

\begin{figure}[ht!]
\centering
\includegraphics[width=9 cm, height=8.5cm]{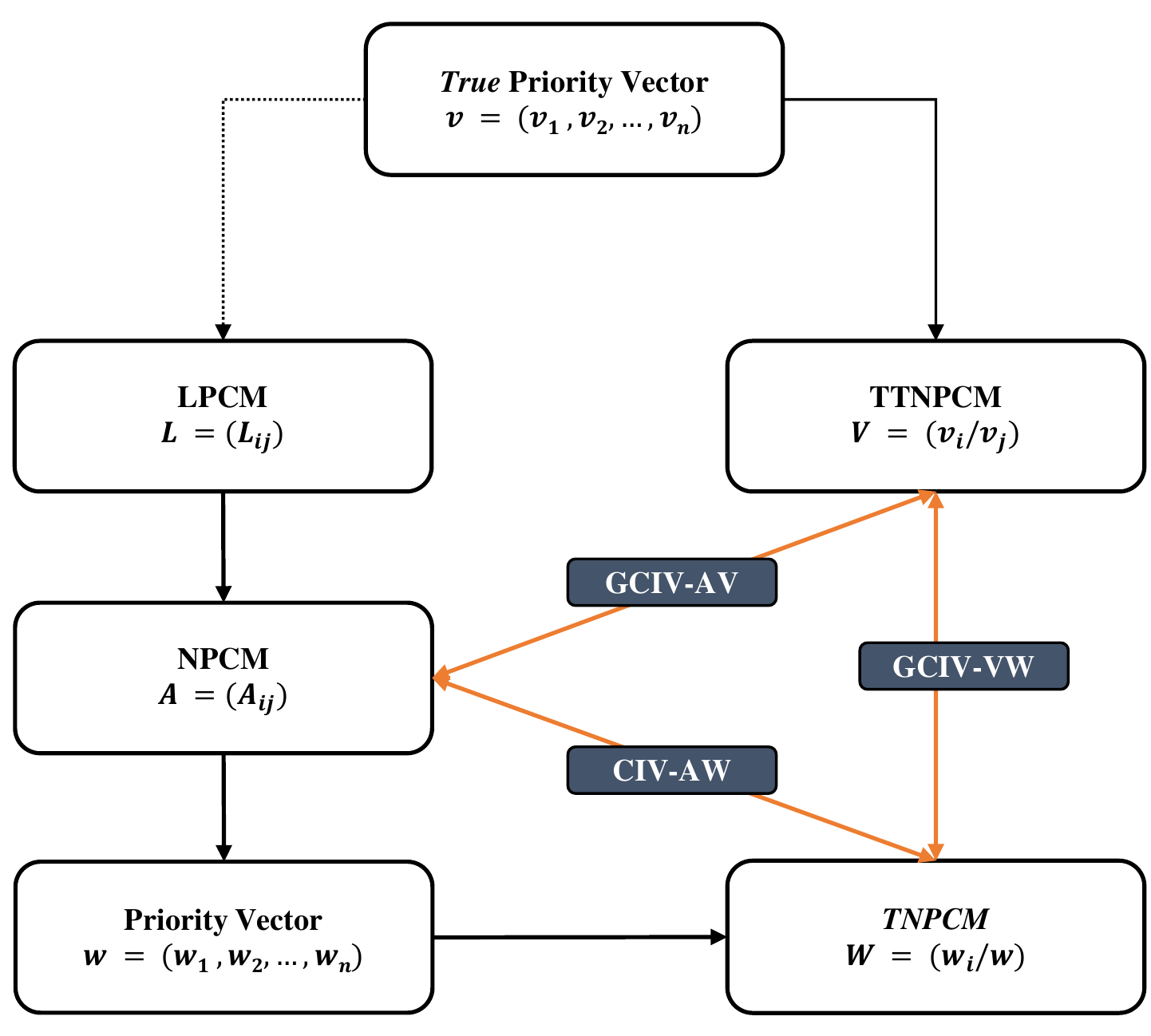}
\caption{AHP \& Compatibility}
\label{GCIVdiagram}
\end{figure}

In the context of compatibility in AHP, there are three possible GCIVs that can be considered. One of them corresponds to the CIV of Saaty introduced in definition 9 and is represented as GCIV-AW in Figure \ref{GCIVdiagram}. The second one is GCIV-VW which addresses the similarity between matrix $V$ derived from \textit{true} priority vector $v$ and the matrix $W$ derived from \textit{calculated} priority vector $w$. And the third one is GCIV-AV which addresses the similarity between matrix $A$ elicited from the decision maker and matrix $V$ derived from \textit{true} priority vector $v$. In a perfectly consistent setting, all these GCIVs would be equal to one. The following sub-section provides an illustrative example for all the definitions presented above. 

    \subsection{Illustrative Example}
    \label{IE_Prelimanries}
    Let's assume that five criteria are to be evaluated and the true priority vector for these five criteria is given as $v=[0.40, 0.30, 0.20, 0.05, 0.05]$. Note that in most cases, this true priority vector will be unknown and the objective of AHP is to assess this vector. Given this true priority vector, let's assume that the linguistic pairwise comparison matrix elicited from the decision maker is as follows:

\begin{equation}\label{IE_LPCM}
L =
\begin{pmatrix}
S9 & S10 & S11 & S15 & S16\\
S8 & S9 & S10 & S13 & S14\\
S7 & S8 & S9 & S14 & S13\\
S3 & S5 & S4 & S9 & S12\\
S2 & S4 & S5 & S6 & S9
\end{pmatrix}    
\end{equation}

Matrix L can be converted into an NPCM using a scale function $f_{(scale)}$ among those that are represented in Table \ref{NumericalScales}. Given a Saaty scale (1 - 9), Matrix L would be converted to NPCM such that $A = f_{(Saaty)}(L)$ as follows: 

\begin{equation}\label{IE_NPCM}
A =
\begin{pmatrix}
1   & 2    & 3     & 7      & 8 \\
1/2 & 1    & 2     & 5      & 6 \\
1/3 & 1/2  & 1     & 6      & 5 \\
1/7 & 1/5  & 1/6   & 1      & 4 \\
1/8 & 1/6  & 1/5   & 1/4    & 1
\end{pmatrix}    
\end{equation}

As proposed by Thomas L. \cite{saaty1977scaling}, \textit{calculated} priority vector is the eigenvector corresponding to the maximum eigenvalue $(\lambda_{max}=5.3436)$ of $A$. This vector is calculated as $w=[0.4329, 0.2671, 0.1975, 0.0669, 0.0356]$. Also, the $CI$ of the matrix $A$ presented in definition 5 is calculated as $\frac{(5.3436-5)}{(5-1)} = 0.0859$. Furthermore, $CR$ presented in definition 6 can be calculated as $\frac{0.0859}{1.12} = 0.0767$, whereas the value of Random Index (R.I) is 1.12 \cite{saaty1977scaling}.  

From the \textit{calculated} priority vector $w=[0.4329, 0.2671, 0.1975, 0.0669, 0.0356]$, TNPCM $W$ presented in definition 7 is given as follows; 

\begin{equation}\label{IE_TNPCM}
W =
\begin{pmatrix}
1.00 &	1.62 &	2.19 &	6.47 &	12.14 \\
0.62 & 	1.00 &	1.35 &	3.99 &	7.49 \\
0.46 &	0.74 &	1.00 &	2.95 &	5.54 \\
0.15 &	0.25 &	0.34 &	1.00 &	1.88 \\
0.08 &	0.13 &	0.18 &	0.53 &	1.00
\end{pmatrix}    
\end{equation}

Furthermore, from the \textit{true} priority vector $v=[0.40, 0.30, 0.20, 0.05, 0.05]$, TTNPCM $V$ presented in definition 8 is given as follows; 

\begin{equation}\label{IE_TTNPCM}
V =
\begin{pmatrix}
1.00 &	1.33 &	2.00 &	8.00 &	8.00 \\
0.75 &	1.00 &	1.50 &	6.00 &	6.00 \\
0.50 &	0.67 &	1.00 &	4.00 &	4.00 \\ 
0.13 &	0.17 &	0.25 &	1.00 &	1.00 \\
0.13 &	0.17 &	0.25 &	1.00 &	1.00
\end{pmatrix}    
\end{equation}

From definitions 9 \& 10, the similarity between any pair of matrices can be measured through a GCIV. In Figure \ref{GCIVdiagram}, various measures of GCIV are presented. These values are calculated as GCIV-AV $= 1.1174$, GCIV-AW $= 1.0687$ and GCIV-VW $ = 1.0443$.

\section{Heuristic for Individualization of Numerical Scale}
\label{Heuristic}
As discussed earlier, a fixed generic scale to convert linguistic labels into numbers diminishes the true meaning of the preferences of the decision maker. Scale individualization is one approach through which this issue can be addressed. Most of the previous studies (\cite{dong2013numerical}; \cite{zhou2018analytic}) on scale individualization focus on reducing the inconsistency of the numerical pairwise comparison matrix. However, efforts to reduce inconsistency in a pairwise comparison matrix can distort the true meaning of the preferences in such a way that it no longer represents decision-maker preferences. A distinction should be made between the consistency of the preferences and the validity of the underlying decision process. Improving the consistency of an NPCM does not necessarily improve the validity of the results and thus consistency improving methods could be misleading \citep{saaty2007invalidity}.  

\begin{figure}[ht!]
\centering
\includegraphics[width=4.6cm, height=10cm]{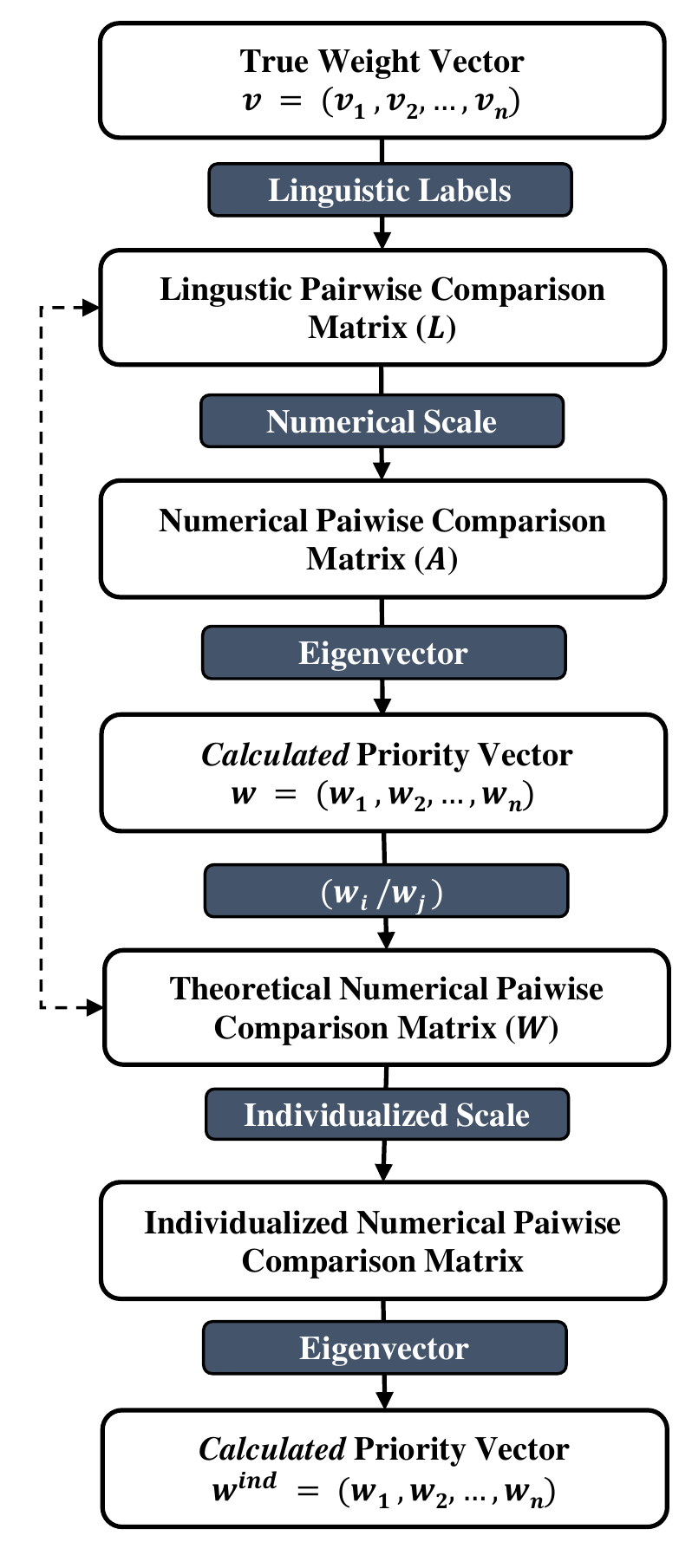}
\caption{A Basic Algorithm for Generating Individualized Numerical Scale (BAGINS)}
\label{Figure_IE}
\end{figure}

We utilize the special structure of pairwise comparison matrices to further extend the process illustrated in Figure~\ref{AHP} and propose a Basic Algorithm for Generating an Individualized Numerical Scale (BAGINS) to generate an individualized numerical scale. While comparing two objects, participants can accurately provide the rank or the order relationship, but capturing the degree of certitude is difficult. The novel heuristic proposed in this study measure this degree of certitude by constructing an individualized numerical scale during the process of estimating priority vectors from pairwise comparison matrices. This heuristic is presented in Algorithm \ref{Heuristic_Ind} and graphically illustrated in Figure~\ref{Figure_IE}. 

\begin{algorithm}[ht!]
\caption{Basic Algorithm for Generating Individualized Numerical Scale (BAGINS)}\label{Heuristic_Ind}
\begin{algorithmic}[1]
\State \textit{Let} $S = (S_k \mid k = 1,2,...,m )$
\State \textit{Elicit} $L = (l_{ij})_{n\times n}$ \textit{where} $l_{ij} \in S$
\State \textit{Construct} $A = (a_{ij})_{n \times n}$ \textit{such that} $A = f_{(Saaty)}(L)$
\State \textit{Calculate} $w$ \textit{such that} $A \times w = \lambda_{max} \times w$
\State \textit{Construct} $W = (w_{ij})_{n \times n}$ \textit{where} $w_{ij}=\frac{w_i}{w_j}$

\State \textit{Initialize} $f_{(ind)}^{k}=0$ \textit{for} $k = 1,2,...,m$  

\State $f_{(ind)}^{\frac{m+1}{2}}=1$

\For {$k = \frac{m+1}{2}+1,...,m$}
\State \textit{Initialize} $T_k=\{ \}$

\For {$\forall i,j$ of $L$}
  \If{$l_{ij}=S_k$}
    \State $T_k=T_k \cup (i,j)$
   \EndIf
\EndFor
\State $f_{(ind)}^{k} = max \Biggl\{\frac {\mathlarger{\mathlarger{\sum}}\limits_{(i,j)\in T_k} w_{ij}}{\mid T_k \mid}\ ,f_{(ind)}^{k-1}\Biggl\}$ $
\forall (i,j) \in T_k$ 
\EndFor

\For {$k = 1, 2, ...,\frac{m-1}{2}$}
\State $f_{(ind)}^{k} = \Bigl( f_{(ind)}^{\frac{m+1}{2}+k} \Bigl) ^{-1}$
\EndFor

\State $f_{(ind)} \rightarrow$ \textit{Individualized Scale} 
\end{algorithmic}
\end{algorithm}

The Algorithm \ref{Heuristic_Ind} takes preferences in the form of linguistic labels as input and generates a personalized numerical scale $f_{(ind)}^k$ for $k = 1,2,...,m$. $S$ is a tuple representing linguistic labels and $L$ is elicited such that $l_{ij} \in S$ for $i,j = 1,2,...,n $. As part of the process, $L$ is converted to $A$ using the Saaty scale so that $A = f_{(Saaty)}(L)$. After calculating the eigenvector $w$ corresponding to the maximum eigenvalue ($\lambda_{max}$) of $A$, a theoretical pairwise comparison matrix $W$ is constructed. Recall that in the Saaty scale, we have 17 linguistic labels, where $S_9$ corresponds to ``equals", i.e., $f_{(Saaty)}^9=1$. Algorithm \ref{Heuristic_Ind} first determines an individualized scale for linguistic labels $S_9$ to $S_{17}$. For the remaining linguistic labels (i.e., $S_1$ to $S_8$), the reciprocal property is observed. Line 8 sets the linguistic label corresponding to ``equal" to have a value of 1. From lines 9 to 12, we determine the $(i,j)$ pairs, where the entry of the $L$ matrix is $S_k$ or $S_{(m-k+1)}$ (i.e., the reciprocal counterpart of $S_k$) and construct a set of these ordered pairs referred to as $T_k$. Line 13 averages the values of the $(i,j)$ entries of the matrix $W$ for all ordered pairs that are in set $T_k$. The maximum operator in this line ensures the monotonicity of the scale. Lines 14 and 15 ensure that the scale is reciprocal.    

    \subsection{Illustrative Example}
    \label{IE_Heuristc}
    Recall the illustrative example presented in subsection \ref{IE_Prelimanries}, a hypothetical linguistic pairwise comparison matrix elicited from the decision maker (i.e., matrix L) is provided by Equation \ref{IE_LPCM}. Suppose that the numerical counterparts of the linguistic labels that the decision maker utilizes while providing the preferences are represented by the inherent scale which is presented in Table \ref{NumericalScalesThisStudy}. The corresponding $A$ matrix constructed from the Saaty scale is given in Equation \ref{IE_NPCM}. The eigenvector corresponding to the maximum eigenvalue of $A$ referred to as the \textit{calculated} priority vector is determined as $w=[0.4329, 0.2671, 0.1975, 0.0669, 0.0356]$ and the theoretical pairwise comparison matrix $W$ constructed from this \textit{calculated} priority vector is given by Equation \ref{IE_TNPCM}. This theoretical matrix $W$ leads to the construction of individualized scales i.e., the numerical counterpart of the linguistic labels ($S_i$).

For example, $l_{12}$ and $l_{23}$ corresponds to $S_{10}$ in matrix $L$. The corresponding numbers in matrix $W$ ($w_{12}$ and $w_{23}$) are $1.62$ and $1.35$. Therefore, the average of these two numbers is determined as the numerical counterpart of the linguistic label $S_{10}$ i.e., $S_{10} = 1.49$. The complete individualized numerical scale obtained from this process is provided in Table \ref{NumericalScalesThisStudy}. Using this scale, a new individualized NPCM is determined as follows: 

\begin{equation}\label{IE_NPCM_IND}
A_{ind} =
\begin{pmatrix}
1.00 &	1.49 &	2.19 &	6.47 &	12.14 \\
0.67 &	1.00 &	1.49 &	4.76 &	5.22 \\
0.46 &	0.67 &	1.00 &	5.22 &	4.76 \\
0.15 &	0.21 &	0.19 &	1.00 &	2.19 \\
0.08 &	0.19 &	0.21 &	0.46 &	1.00
\end{pmatrix}    
\end{equation}

The eigenvector corresponding to the maximum eigenvalue of the matrix $A_{ind}$ is now referred to as the \textit{calculated} priority vector. This vector is calculated as $w=[0.4215, 0.2665, 0.2130, 0.0603, 0.0387]$. Various GCIV measures that are presented in Section \ref{Preliminaries} will be used as the performance metric to compare different scales. For this illustrative example, the GCIV measures are tabulated in Table \ref{CIV_IE}.  

\begin{table}[ht!]
\small
\centering
\caption{Generalized Compatibility Index Values} 
\vspace*{-3mm}
\label{CIV_IE}
\begin{tabular}{lccc}   
\hline
Scales/GCIV & GCIV-AV & GCIV-AW & GCIV-VW \\    
\hline
Saaty scale & $1.1174$ & $1.0687$ & $1.0443$ \\
Inherent scale & $1.0501$ & $1.0272$ & $1.0229$ \\
Individualized & $1.0426$ & $1.0175$ & $1.0245$ \\

\hline
\end{tabular}
\end{table}

\begin{table}[ht!]
\small
\centering
\caption{Numerical scales used to construct NPCM} 
\vspace*{-3mm}
\label{NumericalScalesThisStudy}
\begin{tabular}{lccc}   
\hline
Linguistic Label & Saaty Scale & Inherent Scale & Individualized Scale \\    
\hline
$S_1=$ extremely less important & $0.11$ & $0.08$ & $0.08$ \\
$S_2=$ very, very strongly less important & $0.13$ & $0.11$ & $0.08$ \\
$S_3=$ demonstratedly less important & $0.14$ & $0.13$ & $0.15$ \\
$S_4=$ strongly plus less important & $0.17$ & $0.17$ & $0.19$ \\
$S_5=$ strongly less important & $0.20$ & $0.20$ & $0.21$ \\
$S_6=$ moderately plus less important & $0.25$ & $0.40$ & $0.46$ \\
$S_7=$ moderately less important & $0.33$ & $0.50$ & $0.46$ \\
$S_8=$ weakly less important & $0.50$ & $0.67$ & $0.67$ \\
$S_9=$ equally important & $1.00$ & $1.00$ & $1.00$ \\
$S_{10}=$ weakly more important & $2.00$ & $1.50$ & $1.49$ \\
$S_{11}=$ moderately more important & $3.00$ & $2.00$ & $2.19$ \\
$S_{12}=$ moderately plus more important & $4.00$ & $2.50$ & $2.19$ \\
$S_{13}=$ strongly more important & $5.00$ & $5.00$ & $4.76$ \\
$S_{14}=$ strongly plus more important & $6.00$ & $6.00$ & $5.22$ \\
$S_{15}=$ demonstratedly more important & $7.00$ & $8.00$ & $6.47$ \\
$S_{16}=$ very; very strongly more important & $8.00$ & $9.00$ & $12.14$ \\
$S_{17}=$ extremely more important & $9.00$ & $12.00$ & $12.14$ \\

\hline
\end{tabular}
\end{table}

    \subsection{Motivation Behind BAGINS}
    \label{MotivationofBAGINS}
    Earlier research on scale individualization (\cite{dong2013numerical}, \cite{zhou2018analytic}) aims to reduce the inconsistency of the numerical pairwise comparison matrix where the focus is on deciding the individualized scale that enhances the transitivity relations in the matrix $A$, i.e., $a_{ij} \times a_{jk} =a_{ik}$ $\forall i,j,k$. BAGINS takes an alternative approach where the objective is choosing the individualized scales with the aim that the compatibility between matrix $A$ and matrix $W$ (recall Equation \ref{CIV}) is improved so that:

\begin{equation}\label{CIV_sum_n_square}
\sum_{i=1}^{n} \sum_{j=1}^{n} a_{ij}.\frac{w_j}{w_i}={n^2}
\end{equation}

Note that the sum in Equation \ref{CIV_sum_n_square} holds \textit{iff} $a_{ij}= \frac{w_i}{w_j}$ \citep{saaty1994ratio}, i.e., GCIV-AW=1. The proof is simple when one considers that due to the reciprocal structure of the left hand side term in  Equation \ref{CIV_sum_n_square}, it can be represented as $\frac{n^2}{2}$ pairs of terms of the convex form $x+\frac{1}{x}$ each of which has a minimum value of $2$. Therefore, the minimum value of the double summation is attained when $a_{ij}= \frac{w_i}{w_j}$, i.e., the compatibility is equal to $1$. 

As a result of this observation, BAGINS targets a proxy optimization problem that aims to minimize the following objective function: 

\begin{equation}\label{BAGINS_Objective}
\min \sum_{i=1}^{n} \sum_{j=1}^{n} (a_{ij}-\frac{w_i}{w_j})^2
\end{equation}

Equation \ref{BAGINS_Objective} aims to minimize the sum of the squared deviations between the $a_{ij}$ and $\frac{w_i}{w_j}$, which in turn improves the compatibility between matrix $A$ and matrix $W$. Note that the sample mean is the value that minimizes the sum of the squared deviations. This is taken into consideration in BAGINS at Line 13 which \textit{averages} the values of the $(i,j)$ entries of the matrix $W$ for all ordered pairs that are in set $T_k$.

Targeting the compatibility as opposed to the transitivity is what sets BAGINS apart from the previous scale individualization algorithms. In that aspect, BAGINS objective is similar to the objective of the LLSM \citep{crawford1985note} algorithm which was proposed as one of the earliest alternatives to Saaty's eigenvector approach for deriving the priority vector. The objective of LLSM is as follows:

\begin{equation}\label{LLSM_Objective}
\min \sum_{i=1}^{n} \sum_{j=1}^{n} [\ln a_{ij}-(\ln {w_i} - \ln {w_j})]^2
\end{equation}

LLSM's decision variables are the weights (i.e., $w_i$ $\forall i$) and assume Saaty's fixed scale as the gold standard while converting the linguistic pairwise comparison matrix into the numerical pairwise comparison matrix (i.e., determining the $a_{ij}$). On the other hand, BAGINS objective is determining the individualized scale that will be used to determine the numerical pairwise comparison matrix (i.e., determining the $a_{ij}$) so that the compatibility of the matrix $A$ and the TNPCM derived from it (i.e., $W$) is minimized. That is to say, LLSM determines the priority vector and assumes the fixed Saaty scale, while BAGINS determines the individualized scale and assumes the priority vectors determined by Saaty's eigenvector approach.

\section{Research Methodology}
\label{ResearchMethodology}
In this study, we will evaluate the performance of three methods based on a synthetically created \textit{Numerical Dataset} as well as two \textit{Empirical Datasets}. The methods and the corresponding nomenclature used in this research for each of the three methods are tabulated in Table~\ref{Methods}. 

\begin{table}[ht!]
\small
\centering
\caption{Nomenclature of methods}
\vspace*{-3mm}
\label{Methods}
\begin{tabular}{ll}   \hline
Method & Abbreviation  \\ \hline
Original Method based on linear scale proposed by Thomas L. \cite{saaty1977scaling} & Saaty \\
Individualization using the heuristic proposed in this study & BAGINS \\
Individualization based on Nonlinear Programming (NLP) model proposed by \cite{dong2013numerical} & NLP \\
 \hline
 \end{tabular}
\end{table}

The details of the numerical and empirical studies are provided in the following subsections.

    \subsection{Numerical Dataset}
    \label{EmpiricalDataset}
    A numerical study is conducted through a dataset of linguistic pairwise comparison matrices. For this purpose, a methodology is developed that imitates the process of an expert providing pairwise comparisons. In practice, pairwise comparisons are elicited from the experts in the form of linguistic labels. It is assumed that during this elicitation process, they utilize a certain weight vector, an inherent scale (e.g., similar to the one provided in Table \ref{NumericalScalesThisStudy}) and provide preferences in the form of pairwise comparisons. Due to various reasons such as the inconsistency of the expert, the vagueness associated with the natural language, and the dubious nature of the scale used to quantify these linguistic labels, the resulting NPCM is an inconsistent matrix. Therefore, in order to mimic this elicitation process, which also includes inconsistency, a numerical dataset is generated which consists of NPCMs of four different matrix sizes i.e., $n = 3, 7, 11, 15$ and three levels of inconsistencies i.e., $\textit{Low, Medium, High}$. 

A random normalized vector $w_1, w_2,\cdots,w_n$ is generated that corresponds to the weight vector of an expert evaluating $n$ criteria. Using this random weight vector and Equation~\ref{cMatrix}, a fully consistent pairwise comparison matrix is generated.   

\begin{equation}
W =
 \begin{pmatrix}
  w_{1}/w_{1} & w_{1}/w_{2} & \cdots & w_{1}/w_{n} \\
  w_{2}/w_{1} & w_{2}/w_{2} & \cdots & w_{2}/w_{n} \\
  \vdots  & \vdots  & \ddots & \vdots  \\
  w_{n}/w_{1} & w_{n}/w_{2} & \cdots & w_{n}/w_{n}
 \end{pmatrix}
 \label{cMatrix}
\end{equation}

As previously stated, pairwise comparison matrices in real life applications almost always consist of inherent inconsistency. Therefore, to represent real elicitation process, inconsistency is synthetically added into pairwise comparison matrices using a parameter $\beta \in \left\{0, 0.2, 0.4, 0.6, 0.8, 1.0\right\}$. For each number in the matrix $a_{ij}$ representing pairwise comparison, an interval $[a\;b]$ is created, such that $a=w_i/w_j-\beta \times w_i/w_j$ and $b=w_i/w_j+\beta \times w_i/w_j$. From this interval $[a\;b]$, a number $x_{ij}$ is randomly chosen and replaced with the corresponding $a_{ij}$ in the matrix. The reciprocal property of the pairwise comparison matrices is preserved during this process of adding inconsistency. 

Note that due to the randomness in the methodology employed to incorporate inconsistency, a numerically generated pairwise comparison matrix can have a different level of inconsistency than initially intended through the $\beta$ parameter. In order to address this issue, the real inconsistency of matrices is calculated through $CR = CI/RI$, and then this $CR$ measure is used for the classification of matrices on different levels of inconsistency. A $CR$ value between $0-0.03$ corresponds to \textit{low level} inconsistent matrices. Similarly, a $CR$ value between $0.03 - 0.06$ corresponds to \textit{medium level} inconsistent matrices. While all matrices with $CR$ between $0.06 - 0.1$ are classified as a \textit{high level} inconsistent matrices. Any matrix with a $CR$ value $\geq$0.1 is regarded as not sufficiently consistent \citep{saaty1977scaling} and therefore such matrices are discarded from the data set. 

Next, an inverse scale function and Saaty scale of 1-9 is used to transform this data set of inconsistent NPCM into LPCM. Weight vectors generated at the start of this process are considered as true weights and numerically generated LPCMs are considered as preferences elicited from the expert. This procedure is summarized in Algorithm 1.   

\begin{algorithm}[ht!]
\caption{Generate Synthetic Pairwise Comparison Matrices}\label{euclid}
\begin{algorithmic}[1]
\State $\textit{n=3}$
\While {$\; n\leq 15$}
\State $\textit{True Weights}\gets \textit{Generate n Random Weight which sums upto 1}$
\State $\textit{W} = \{ w_i/w_j \}$
\State $\beta=0$
\While {$\; \beta\leq 1$}
\State $\textit{NPCM = Add Inconsistency(NPCM)} $
\State $LPCM = f^{-1}(NPCM)$
\State $\beta = \beta+0.2$
\EndWhile
\State $n = n+4$
\EndWhile
\State $Final Dataset = sort(LPCM)$
\end{algorithmic}
\end{algorithm}

The initial data set generated through Algorithm 1 is composed of 4,800 pairwise comparison matrices (i.e., $4\; \text{number of matrix sizes}\times 6 \; \text{number of $\beta$ parameter} \times 200 \; \text{replications}$). However, due to the uneven distribution of the \textit{low-level} and \textit{high-level} inconsistent matrices, the dataset is reclassified based on the $CR$ value (instead of $\beta$). After reclassification, the initial data set of 4,800 matrices is reduced to a total of 900 matrices (i.e., $4\; \text{number of matrix sizes}\times 3 \; \text{consistency levels} \times 75 \; \text{replications}$) which are \textit{evenly} distributed on all experimental parameters. For each combination of $n$ and $CR$, 75 replications were made and thus \textit{final data set} is composed of 900 $(=4\times3\times75)$ matrices. 

    \subsection{Empirical Dataset}
    \label{NumericalDataset}
\begin{figure}[ht!]
\centering
\includegraphics[width=5cm, height=5cm]{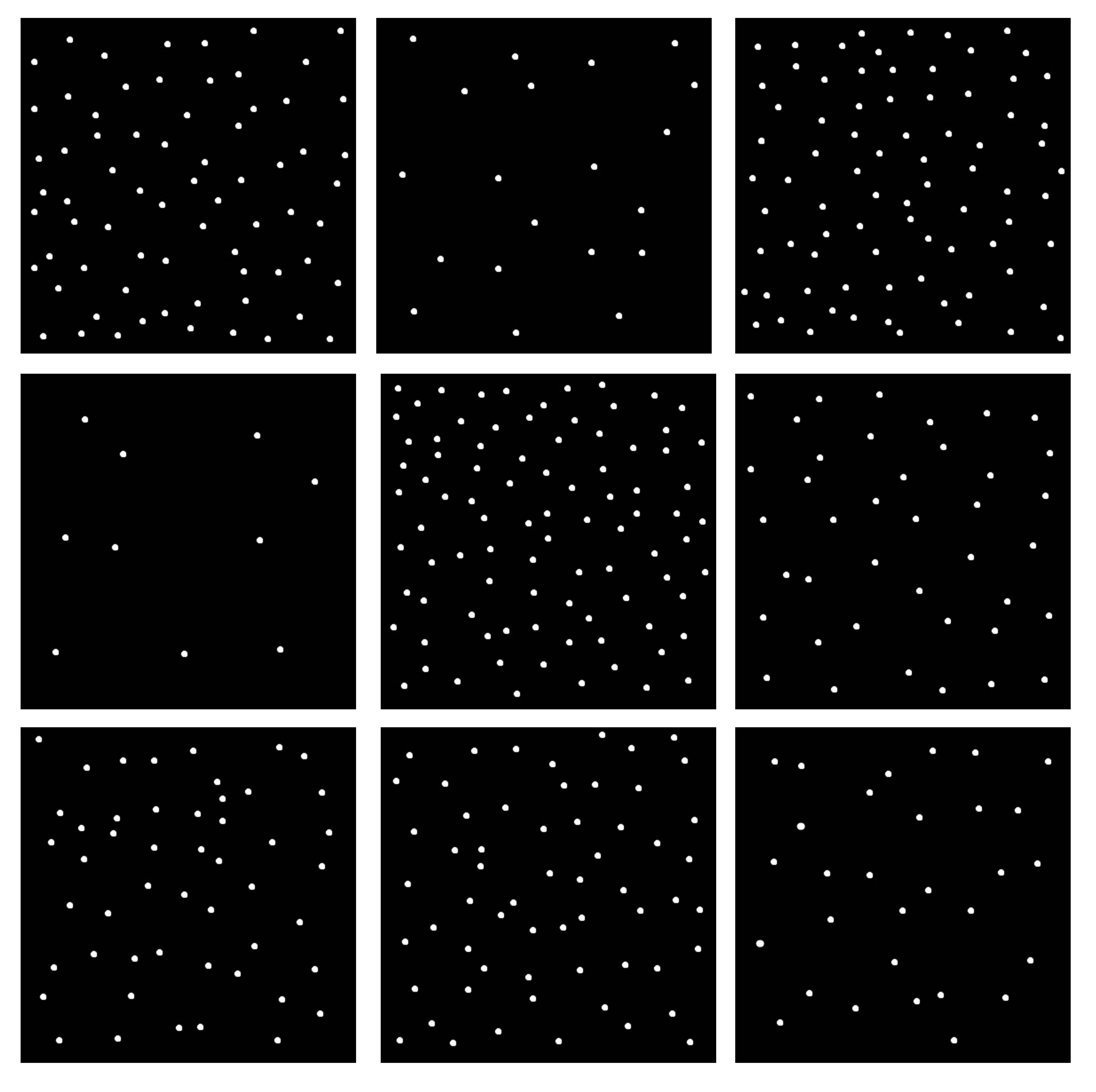}
\caption{Visual Experiment to seek pairwise comparisons of different densities}
\label{DensityExperiment}
\end{figure}

    We conducted two experiments for which true priority vectors can be measured due to the inherent natural scales in the experiments. The first experiment deals with visual observations in which participants were shown two images at a time and asked to provide pairwise comparisons in the form of linguistic labels such as \textit{``Extremely Dense''}, \textit{``Moderately Dense''} etc. For this experiment, nine different images (Figure~\ref{DensityExperiment}) were employed. In the second experiment consisting of nine different bottles, participants were asked to weigh two bottles at a time and provide their preferences in the form of linguistic labels such as \textit{``Extremely Heavy''}, \textit{``Moderately Heavy''} etc. Bottles were covered with black paint so that visual observations did not impact their preferences. 

The participants in both studies were undergraduate students enrolled at Sabanci University, Istanbul, Turkey. In the visual experiment, there were 164 participants; in the mass experiment, 154 participants. For both experiments, a total of 36 comparisons were done by each student and there was no specific ordering of the comparisons. For their participation in the experiments, students received 2 bonus points for their course grades. To ensure their focus during the experiments, an additional 1 bonus point was promised for the top performing participants in terms of their consistency index value.  

Both experiments were approved by the university ethics committee and written consent of the students was obtained prior to participation in the study. The average time required for the visual experiment and the mass experiment was approximately 15 minutes and 20 minutes, respectively per participant. Since both experiments had an underlying natural scale (i.e., for the visual experiment, the number of dots; for the mass experiment, grams), it was possible to assess the \textit{true weights} and corresponding weight vector for both experiments as tabulated in Table~\ref{TrueWeights}. 

\begin{table}[ht!]
\small 
\centering
\caption{True normalized weight vector for visual and mass experiment}
\vspace*{-3mm}
\label{TrueWeights}
\begin{tabular}{ccc}   \hline
Number of Dots  & Mass of Bottles (Grams) & Weight Vector \\  \hline
10 & 50 & 0.0222 \\
20 & 100 & 0.0444 \\ 
30 & 150 & 0.0667 \\ 
40 & 200 & 0.0889 \\ 
50 & 250 & 0.1111 \\ 
60 & 300 & 0.1333 \\ 
70 & 350 & 0.1556 \\ 
80 & 400 & 0.1778 \\ 
90 & 450 & 0.2000 \\ 
 \hline
 \end{tabular}
\end{table}

For researchers and practitioners working in similar domains, the  numerical and empirical datasets of pairwise comparison matrices along with necessary Matlab codes and documentation are made available online at \citep{ahmedfaranGitHub}.  

\section{Results and Discussion}
\label{ResultsAndDiscussion}
In this section, we present and discuss the results of both the numerical and empirical studies. We considered all three Generic Compatibility Index Values that were introduced in Section 2 (i.e., GCIV-AW, GCIV-VW, GCIV-AV) as the performance metrics of the methods that are considered in this study. Recall that GCIV-AW is the original CIV, which was introduced by Saaty and interrelated with the CI metric as we have discussed earlier. On the other hand, both for the numerical study and the empirical studies we also have the true priority vectors. Therefore, it is also possible to determine the GCIV-VW and GCIV-AV while comparing the performance of the three methods that are tabulated in Table \ref{Methods}.

    \subsection{Results of the Numerical Study}
    \label{NumericalAnalysis}
    Table \ref{GCIV_AW_VW_AV_n} tabulates the mean GCIV-AW, GCIV-VW and GCIV-AV for different matrix sizes. Results demonstrate that the mean CIV between the priority vector and the corresponding NPCM (i.e., GCIV-AW), is minimum for the NLP for all matrix sizes (i.e., $n=3, 7, 11, \text{ and } 15$). The results from the LSD post hoc test (Table \ref{LSD_GCIV_AW_VW_AV_n}) indicate that these differences are statistically significant $(P\leq 0.05)$. This result is not surprising due to the objective function of the NLP. The NPCMs constructed from the LP model is highly consistent and therefore the priority vector calculated from these matrices would easily yield minimum CIV when compared with the corresponding NPCM. On the other hand, BAGINS also outperforms Saaty in this metric and the difference is statistically significant for all matrix sizes.   

Mean GCIV-VW, which addresses the deviation of the calculated priority vector from the true priority vector is tabulated in Table \ref{GCIV_AW_VW_AV_n}. For the smallest matrix size (i.e., $n=3$), all of the three methods perform similarly and there are no statistically significant differences among them (Table \ref{LSD_GCIV_AW_VW_AV_n}). However, for larger matrix sizes (i.e., $n=7, 11, \text{ and } 15 $), the NLP method is significantly outperformed by the other two methods (i.e., Saaty and BAGINS). There isn't a statistically significant difference between Saaty and BAGINS when $n=7$. Saaty significantly outperforms BAGINS only for large matrix sizes (i.e., $n=11$ and $15$). That is to say, in terms of the GCIV-VW metric, Saaty is the best performing method and NLP is the worst performing method for different matrix sizes. 

The third comparison is in terms of the mean GCIV-AV which demonstrates the deviation between the true priority vector and the matrix constructed from a particular scale. Again for the smallest matrix size (i.e., $n=3$) there is no statistically significant difference among the three methods. Table \ref{LSD_GCIV_AW_VW_AV_n} reveals that as the size of the matrix is increased (i.e., $n=7, 11, \text{ and } 15 $), the BAGINS outperforms NLP and the differences in the mean GCIV-AV are statistically significant $(P\leq 0.05)$. BAGINS also outperforms Saaty statistically significantly for $n= 11 \text{ and } 15 $).   

\begin{table}[ht!]
\small
\centering
\caption{GCIV-AW, GCIV-VW, GCIV-AV for different matrix sizes. (\textbf{\textit{Bold}} \textit{figures represents the best results})}
\vspace*{-3mm}
\label{GCIV_AW_VW_AV_n}
\begin{tabular}{cllcccc}  
\hline
Performance Metric & Method & N & Mean ($n=3$) & Mean ($n=7$) & Mean ($n=11$) & Mean ($n=15$) \\ 
\hline
       & NLP    & 225  & \textbf{1.00217} & \textbf{1.00789} & \textbf{1.01048} & \textbf{1.01411} \\
GCIV-AW & Saaty  & 225  & 1.01936 & 1.05830 & 1.07163 & 1.07942 \\
       & BAGINS    & 225  & 1.00697 & 1.02591 & 1.02945 & 1.03133 \\
\hline
         & NLP   & 225  & 1.07531 & 1.07959 & 1.08061 & 1.08200 \\
GCIV-VW  & Saaty & 225  & 1.05852 & \textbf{1.02675} & \textbf{1.01916} & \textbf{1.01597} \\
         & BAGINS   & 225  & \textbf{1.05850} & 1.03210 & 1.03015 & 1.03154 \\
\hline
        &  NLP   & 225 & 1.07904 & 1.08796 & 1.09188 & 1.09685 \\
GCIV-AV &  Saaty & 225 & 1.08010 & 1.08579 & 1.09105 & 1.09529 \\
        &  BAGINS   & 225 & \textbf{1.06602} & \textbf{1.05869} &\textbf{ 1.06041} & \textbf{1.06348} \\
\hline
 \end{tabular}
\end{table}

\begin{table}[ht!]
\footnotesize
\caption{Post Hoc LSD Test for different matrix sizes.}
\vspace*{-3mm}
\label{LSD_GCIV_AW_VW_AV_n}
\begin{tabularx}{17cm}{m{1.5cm}m{0.8cm}m{0.8cm}m{1.6cm}m{0.75cm}m{1.6cm}m{0.75cm}m{1.6cm}m{0.75cm}m{1.6cm}m{0.75cm}}  
\hline
Performance Metric & Method (I) & Method (J) & Mean Diff. (I-J) \textit{$n=3$} & Sig. & Mean Diff. (I-J) \textit{$n=7$} & Sig. & Mean Diff. (I-J) \textit{$n=11$} & Sig. & Mean Diff. (I-J) \textit{$n=15$} & Sig.\\ 
\hline
& NLP     & Saaty & -0.01719 & 0.000 & -0.05041 & 0.000 & -0.06115 & 0.000 & -0.06531 & 0.000 \\
GCIV-AW & NLP     & BAGINS   & -0.00480 & 0.000 & -0.01802 & 0.000 & -0.01896 & 0.000 & -0.01722 & 0.000 \\
& Saaty   & BAGINS   &  0.01238 & 0.000 &  0.03239 & 0.000 &  0.04218 & 0.000 &  0.04809 & 0.000 \\
\hline
& NLP     & Saaty &  0.01679 & 0.246 &   0.05284 & 0.000 &   0.06145 & 0.000 &   0.06603 & 0.000 \\
GCIV-VW & NLP     & BAGINS   &  0.01681 & 0.225 &   0.04749 & 0.000 &   0.05046 & 0.000 &   0.05046 & 0.000 \\
& Saaty   & BAGINS   &  0.00002 & 1.000 &  -0.00534 & 0.095 &  -0.01099 & 0.000 &  -0.01556 & 0.000 \\
\hline
& NLP     & Saaty & -0.00106 & 0.996 &  0.00217 & 0.877 &  0.00083 & 0.980 &  0.00156 & 0.933 \\
GCIV-AV & NLP     & BAGINS   &  0.01301 & 0.462 &  0.02927 & 0.000 &  0.03147 & 0.000 &  0.03338 & 0.000 \\
& Saaty   & BAGINS   &  0.01408 & 0.528 &  0.02711 & 0.095 &  0.03064 & 0.000 &  0.03181 & 0.000 \\

\hline
\end{tabularx}
\end{table}

Table \ref{GCIV_AW_VW_AV_CR} tabulates mean GCIV-AW, GCIV-VW and GCIV-AV at different levels of inconsistencies. Mean CIV between the priority vector and the corresponding NPCM (GVIV-AW) is minimum for NLP at all of the three levels of inconsistency similar to the previous analysis. Table~\ref{LSD_GCIV_AW_VW_AV_CR} indicates that these differences are statistically significant $(P\leq 0.05)$. 

Deviation of the calculated priority vector from the true priority vector (GCIV-VW) is minimum for Saaty and BAGINS and results from Table~\ref{LSD_GCIV_AW_VW_AV_CR} depict that both these methods significantly outperform NLP at different levels on inconsistency. In terms of GCIV-VW, the difference between Saaty and BAGINS is not statistically significant for different levels of inconsistencies.   

For GCIV-AV, which addresses the deviation between the true priority vector and the matrix constructed from a particular scale, the BAGINS method provides better results as compared to the other two methods. At low levels of inconsistency, both BAGINS and Saaty significantly outperform NLP, at medium inconsistency levels, BAGINS significantly outperforms the other two methods, while at higher levels of inconsistency, BAGINS and NLP significantly outperform Saaty. 

\begin{table}[ht!]
\small
\centering
\caption{GCIV-AW, GCIV-VW, GCIV-AV for different levels of inconsistency (\textbf{\textit{Bold}} \textit{figures represents the best results})}
\vspace*{-3mm}
\label{GCIV_AW_VW_AV_CR}
\begin{tabular}{cllccc}  
\hline
Performance Metric & Method & N & Mean ($CR=$ \textit{low}) & Mean ($CR=$ \textit{medium}) & Mean ($CR=$\textit{high}) \\ 
\hline
        & NLP     & 225   & \textbf{1.00219} & \textbf{1.00877} & \textbf{1.01503} \\
GCIV-AW &Saaty   & 225   & 1.01597 & 1.05621 & 1.09935 \\
        & BAGINS     & 225   & 1.01115 & 1.02660 & 1.03250 \\
\hline
        & NLP     & 225   & 1.07851 & 1.07782 & 1.08180 \\
GCIV-VW &Saaty   & 225   & \textbf{1.01088} & \textbf{1.03183} & \textbf{1.04760} \\
        & BAGINS     & 225   & 1.01347 & 1.04052 & 1.06023 \\
\hline
        & NLP     & 225   & 1.08066 & 1.08817 & 1.09796 \\
GCIV-AV & Saaty   & 225   & 1.02677 & 1.08886 & 1.14854 \\
        & BAGINS     & 225   & \textbf{1.02456} & \textbf{1.06776} & \textbf{1.09412} \\
\hline
 \end{tabular}
\end{table}

\begin{table}[ht!]
\footnotesize
\caption{Post Hoc LSD Test for different levels of inconsistency - (GCIV-AW)}
\vspace*{-3mm}
\label{LSD_GCIV_AW_VW_AV_CR}
\begin{tabularx}{17cm}{m{1.5cm}m{1.2cm}m{1.2cm}m{2.2cm}m{0.75cm}m{2.2cm}m{0.75cm}m{2.2cm}m{0.75cm}}
\hline
Performance Metric & Method (I) & Method (J) & Mean Diff. (I-J) $CR= $\textit{low} & Sig. & Mean Diff. (I-J) $CR= $\textit{medium} & Sig. & Mean Diff. (I-J) $CR= $\textit{high} & Sig. \\ 
\hline
& NLP     & Saaty & -0.01378 & 0.000 & -0.04744 & 0.000 & -0.08431 & 0.000  \\
GCIV-AW & NLP     & BAGINS   & -0.00896 & 0.000 & -0.01783 & 0.000 & -0.01747 & 0.000  \\
& Saaty   & BAGINS   &  0.00483 & 0.000 &  0.02961 & 0.000 &  0.06684 & 0.000  \\
\hline
& NLP     & Saaty &   0.06764 & 0.000 &   0.04599 & 0.000 &   0.03420 & 0.000  \\
GCIV-VW & NLP     & BAGINS   &   0.06505 & 0.000 &   0.03729 & 0.000 &   0.02158 & 0.000  \\
& Saaty   & BAGINS   &  -0.00259 & 0.517 &  -0.00869 & 0.470 &  -0.01263 & 0.079  \\
\hline
& NLP     & Saaty &  0.05389 & 0.000 & -0.00069 & 0.995 & -0.05057 & 0.000  \\
GCIV-AV & NLP     & BAGINS   &  0.05610 & 0.000 &  0.02041 & 0.009 &  0.00384 & 0.683  \\
& Saaty   & BAGINS   &  0.00221 & 0.627 &  0.02110 & 0.014 &  0.05442 & 0.000  \\
\hline
\end{tabularx}
\end{table}

The results from the numerical study demonstrate that the two methods that are based on individualized scales (i.e., NLP\textit{ or} BAGINS) outperforms the conventional Saaty method which is based on fixed scale in terms of GCIV-AW and GCIV-AV metrics. The differences are always statistically significant regardless of the matrix sizes and the levels of inconsistencies in the case of GCIV-AW. On the other hand, for the case of GCIV-AV the results are statistically significant for larger matrix sizes (i.e., $n= 11 \text{ and } 15 $) and when there is \textit{Medium} or \textit{High} inconsistency in the pairwise comparison matrices. 

Among the two methods that are based on individualization NLP outperforms BAGINS for GCIV-AW as expected due to its objective function. However for the other two metrics (i.e., GCIV-VW and GCIV-AV), BAGINS outperforms NLP\textit{ on the average} regardless of the matrix sizes and inconsistency levels. In terms of GCIV-VW, BAGINS outperforms NLP statistically significantly for all cases. On the other hand, for GCIV-AV, there is no statistically significant difference between the algorithms only for the smallest matrix size (i.e., $n= 3$) and \textit{High} level of inconsistency. For the remaining cases, BAGINS outperforms NLP statistically significantly.


    \subsection{Results of the Empirical Study}
    \label{EmpiricalAnalysis}
    Numerical studies based on synthetic data are highly valuable in order to analyze the performance of alternative algorithms for different experimental conditions (e.g., different matrix sizes, different levels of inconsistency). Compared to a numerical study, an empirical study is constrained by the conditions of a particular context, thus, the conclusions. Having said that, even though empirical studies represent a particular instance of a bigger picture, they are \textit{real} and science aims to give meaning to reality. Thus, empirical studies are invaluable for the development of the theory of AHP as indicated by various researchers (e.g., \cite{brunelli2018survey}, \cite{cavallo2019comparing},
 \cite{sato2022inconsistency}).

With this motivation, we wanted to compare the performance of the three alternative methods, namely, NLP, Saaty and BAGINS with two empirical studies (Visual Experiment and Mass Experiment). We again used GCIV-AW, GCIV-VW and GCIV-AV as three performance metrics in the analysis. The means of the performance metrics for both of the experiments are tabulated in Table~\ref{EmpiricalDS_Overall}. The results imply that in both of the experiments the methods that are based on individualized scales outperform Saaty's method which is based on the fixed scale in both of the experiments. The difference between the performance of NLP and Saaty for all three metrics is statistically significant in both experiments. On the other hand, BAGINS outperforms Saaty in terms of GCIV-AW for both of the experiments statistically significantly. For the other two metrics (namely, GCIV-VW and GCIV-AV) the differences are not statistically significant for two methods.

Even though we didn't have the chance to analyze the performance of the algorithms in terms of the matrix sizes due to the design of the experiments, we conducted further analysis in terms of the level of inconsistencies. The results of this analysis are provided in Table~\ref{EmpiricalDS_Visual} for the Visual Experiment and Table~\ref{EmpiricalDS_Mass} for the Mass Experiment. The LSD Post-Hoc Test results are tabulated in Table~\ref{LSD_Visual} and Table~\ref{LSD_Mass} for the Visual and Mass Experiments, respectively.

The results indicate that NLP outperforms the other two algorithms statistically significantly for all three metrics in both of the experiments when the inconsistency levels are \textit{High}. This is also the case when the inconsistency levels are \textit{Medium} with the exception of the GCIV-VW metric for the Visual Experiment (in this case there is no statistically significant difference between NLP and Saaty). On the other hand, when the inconsistency levels are \textit{Low} the results are blurred. NLP outperforms the other two algorithms only for the GCIV-AW metric in both of the experiments. Particularly for the Visual Experiment Saaty outperforms NLP on the average (i.e., the difference is not statistically significant with $(P\leq 0.05)$) and in terms of GCIV-AV metric BAGINS outperforms both algorithms statistically significantly. 

\begin{table}[ht!]
  \centering
  \caption{Descriptive Statistics (\textbf{\textit{Bold}} \textit{figures represents the best results})}
  \vspace*{-3mm}
    \begin{tabular}{p{2cm}cccc}
    \hline
     &  & Visual Experiment &  &  \\
    \hline
    Model & N & GCIV-AW & GCIV-VW & GCIV-AV  \\
    \hline
    NLP   & 164   & \textbf{1.02785} &\textbf{1.12557} & \textbf{1.15548} \\
    Saaty & 164   & 1.07727 & 1.15077 & 1.22999 \\
    BAGINS   & 164   & 1.04858 & 1.16318 & 1.21772 \\
    \hline
         &  & Mass Experiment &  &  \\
    \hline
        NLP   & 154   & \textbf{1.04640} & \textbf{1.09973} & \textbf{1.14823} \\
    Saaty & 154   & 1.10109 & 1.14718 & 1.25588 \\
    BAGINS   & 154   & 1.07269 & 1.14751 & 1.22790 \\
    \hline
    \end{tabular}
  \label{EmpiricalDS_Overall}
\end{table}

\begin{table}[ht!]
  \centering
  \caption{LSD Post-Hoc Test (Visual Experiment)}
    \vspace*{-3mm}
    \begin{tabular}{p{3em}p{3em}cccccc}
     \hline
     Method  & Method   & GCIV-AW &  & GCIV-VW &  & GCIV-AV &  \\
    \hline
     \textbf{(I)} & \textbf{(J)} & \textbf{Mean Diff. (I-J)} & \textbf{Sig.} & \textbf{Mean Diff. (I-J)} & \textbf{Sig.}& \textbf{Mean Diff. (I-J)} & \textbf{Sig.}\\
    \hline
    NLP   & Saaty & -0.04942 & 0.000 & -0.02520 & 0.004 & -0.07452 & 0.000 \\
    NLP   & BAGINS   & -0.02074 & 0.000 & -0.03761 & 0.000 & -0.06225 & 0.000 \\
    Saaty  & BAGINS   & 0.02868 & 0.000 & -0.01240 & 0.415 & 0.01227 & 0.590 \\
    \hline
    \end{tabular}%
  \label{EmpiricalLSDVisual}%
\end{table}%

\begin{table}[ht!]
  \centering
  \caption{LSD Post-Hoc Test (Mass Experiment)}
    \vspace*{-3mm}
    \begin{tabular}{p{3em}p{3em}cccccc}
     \hline
     Method  & Method   & GCIV-AW &  & GCIV-VW &  & GCIV-AV &  \\
    \hline
     \textbf{(I)} & \textbf{(J)} & \textbf{Mean Diff. (I-J)} & \textbf{Sig.} & \textbf{Mean Diff. (I-J)} & \textbf{Sig.}& \textbf{Mean Diff. (I-J)} & \textbf{Sig.}\\
    \hline
    NLP   & Saaty & -0.05469 & 0.000 & -0.04746 & 0.000 & -0.10765 & 0.000 \\
    NLP   & BAGINS   & -0.02629 & 0.000 & -0.04779 & 0.000 & -0.07966 & 0.000 \\
    Saaty  & BAGINS   & 0.02840 & 0.000 & -0.00033 & 0.999 & 0.02799 & 0.060 \\
    \hline
    \end{tabular}
  \label{EmpiricalLSDMass}
\end{table}

\begin{table}[ht!]
  \centering
  \caption{Descriptive Statistics for different levels of inconsistency (Visual Experiment) (\textbf{\textit{Bold}} \textit{figures represents the best results})}
  \vspace*{-3mm}
    \begin{tabular}{cp{4em}cccccc}
    \hline
    Performance Metric & Model   & N    & CR=low    & N & CR=medium  & N & CR=High \\
    \hline
    & NLP   & 19    & \textbf{1.01238} & 94    & \textbf{1.02104} & 33    &\textbf{ 1.03149} \\
    GCIV-AW & Saaty & 19    & 1.03055 & 94    & 1.05783 & 33    & 1.09362 \\
    & BAGINS   & 19    & 1.02173 & 94    & 1.03547 & 33    & 1.05721 \\
    \hline
    & NLP   & 19    & 1.14716 & 94    & \textbf{1.11738} & 33    & \textbf{1.12571} \\
    GCIV-VW     & Saaty & 19    & \textbf{1.11687} & 94    & 1.13615 & 33    & 1.17516 \\
    & BAGINS   & 19    & 1.11996 & 94    & 1.14311 & 33    & 1.19300 \\
    \hline
    & NLP   & 19    & 1.16059 & 94    & \textbf{1.13988} & 33    & \textbf{1.15853} \\
    GCIV-AV & Saaty & 19    & 1.14956 & 94    & 1.19526 & 33    & 1.26969 \\
    & BAGINS   & 19    & \textbf{1.14503} & 94    & 1.18343 & 33    & 1.25630 \\
    \hline
    \end{tabular}%
  \label{EmpiricalDS_Visual}%
\end{table}%

\begin{table}[ht!]
  \centering
  \caption{Descriptive Statistics for different levels of inconsistency (Mass Experiment)(\textbf{\textit{Bold}} \textit{figures represents the best results})}
  \vspace*{-3mm}
    \begin{tabular}{cp{4em}cccccc}
    \hline
    Performance Metric & Model   & N    & CR=low    & N & CR=medium  & N & CR=High \\
    \hline
    & NLP   & 7     & \textbf{1.01343} & 63    & \textbf{1.02753} & 51    & \textbf{1.04322} \\
    GCIV-AW & Saaty & 7     & 1.03447 & 63    & 1.06003 & 51    & 1.09972 \\
    & BAGINS   & 7     & 1.02370 & 63    & 1.04059 & 51    & 1.06762 \\
    \hline
    & NLP   & 7     & \textbf{1.09956} & 63    & \textbf{1.10267} & 51    & \textbf{1.09012} \\
    GCIV-VW & Saaty & 7     & 1.11979 & 63    & 1.13541 & 51    & 1.14604 \\
    & BAGINS   & 7     & 1.12289 & 63    & 1.13120 & 51    & 1.14154 \\
    \hline
    & NLP   & 7     & \textbf{1.11353} & 63    & \textbf{1.13275} & 51    & \textbf{1.13565} \\
    GCIV-AV & Saaty & 7     & 1.15687 & 63    & 1.20062 & 51    & 1.25567 \\
    & BAGINS   & 7     & 1.14974 & 63    & 1.17802 & 51    & 1.21954 \\
    \hline
    \end{tabular}%
  \label{EmpiricalDS_Mass}%
\end{table}%

\begin{table}[ht!]
\small
  \centering
  \caption{LSD Post-Hoc Test for different levels of inconsistency (Visual Experiment)}
    \vspace*{-3mm}
    \begin{tabular}{cp{3em}p{3em}cccccc}
    \hline
    & Method  & Method & CR=low & & CR=medium &  & CR=high &  \\
    \hline
    Performance Metric & (I)   & (J) & Mean Diff. (I-J) & Sig. & Mean Diff. (I-J) & Sig.& Mean Diff. (I-J) & Sig. \\
    \hline
    & NLP   & Saaty & -0.01816 & 0.000 & -0.03679 & 0.000 & -0.06212 & 0.000 \\
    GCIV-AW & NLP   & BAGINS   & -0.00935 & 0.003 & -0.01443 & 0.000 & -0.02572 & 0.000 \\
    & Saaty  & BAGINS   & 0.00882 & 0.000 & 0.02236 & 0.000 & 0.03640 & 0.000 \\
    \hline 
    & NLP   & Saaty & 0.03029 & 0.232 & -0.01877 & 0.134 & -0.04945 & 0.017 \\
    GCIV-VW & NLP   & BAGINS   & 0.02720 & 0.327 & -0.02573 & 0.041 & -0.06729 & 0.004 \\
    & Saaty  & BAGINS   & -0.00309 & 0.986 & -0.00696 & 0.800 & -0.01784 & 0.670 \\
    \hline
    & NLP   & Saaty & 0.01103 & 0.822 & -0.05537 & 0.000 & -0.11116 & 0.000 \\
    GCIV-AV & NLP   & BAGINS   & 0.01556 & 0.700 & -0.04355 & 0.000 & -0.09777 & 0.000 \\
    & Saaty  & BAGINS   & 0.00453 & 0.973 & 0.01183 & 0.538 & 0.01339 & 0.785 \\
    \hline
    \end{tabular}
  \label{LSD_Visual}
\end{table}

\begin{table}[ht!]
\small
  \centering
  \caption{LSD Post-Hoc Test for different levels of inconsistency (Mass Experiment)}
    \vspace*{-3mm}
    \begin{tabular}{cp{3em}p{3em}cccccc}
    \hline
    & Method  & Method & CR=low & & CR=medium &  & CR=high &  \\
    \hline
    Performance Metric & (I)   & (J) & Mean Diff. (I-J) & Sig. & Mean Diff. (I-J) & Sig.& Mean Diff. (I-J) & Sig. \\
    \hline
    & NLP   & Saaty & -0.02104 & 0.000 & -0.03250 & 0.000 & -0.05650 & 0.000 \\
    GCIV-AW & NLP   & BAGINS   & -0.01027 & 0.012 & -0.01306 & 0.000 & -0.02440 & 0.000 \\
    & Saaty  & BAGINS   & 0.01077 & 0.005 & 0.01945 & 0.000 & 0.03210 & 0.000 \\
    \hline 
    & NLP   & Saaty & -0.02023 & 0.700 & -0.03274 & 0.000 & -0.05592 & 0.000 \\
    GCIV-VW & NLP   & BAGINS   & -0.02333 & 0.697 & -0.02853 & 0.004 & -0.05142 & 0.000 \\
    & Saaty  & BAGINS   & -0.00310 & 0.994 & 0.00421 & 0.881 & 0.00450 & 0.926 \\
    \hline
    & NLP   & Saaty & -0.04334 & 0.228 & -0.06787 & 0.000 & -0.12003 & 0.000 \\
    GCIV-AV & NLP   & BAGINS   & -0.03621 & 0.407 & -0.04528 & 0.000 & -0.08389 & 0.000 \\
    & Saaty  & BAGINS   & 0.00713 & 0.969 & 0.02260 & 0.030 & 0.03613 & 0.019 \\
    \hline
    \end{tabular}
  \label{LSD_Mass}
\end{table}

\section{Concluding Remarks and Future Research}
\label{Conclusions}
Most decision-making environments contain qualitative information in the form of verbal phrases and the quantification of such phrases has remained a contentious issue in the literature. With the advance of digital transformation in all aspects of our daily lives, human-machine teaming is becoming \textit{normal}, thus, accurate quantification of such verbal phrases is of critical importance while eliciting judgments (e.g., probability, weights, etc.) from human experts and/or decision-makers. 

Being the most popular method in the realm of MCDM, AHP always received excessive attention from both researchers and practitioners. Since the early days that it was introduced, one of the major controversies regarding to AHP was the \textit{arbitrary scale} that was utilized while transforming the linguistic comparison matrices to the numerical comparison matrices which are used to determine the true priority vectors of the decision-makers. Therefore, many alternative \textit{static} scales are proposed in the literature. Recently, the individualization of the scale that is used in the process has been receiving attention from researchers. In this study, we have introduced BAGINS, a simple, easy-to-learn, and quantitatively less demanding heuristic that performs comparable with other alternatives.

Our contribution in this study is threefold. First of all, we have developed an experimental setup that can be used not only for the problem in our focus (i.e., the individualization of the scales) but also for many other researchers that are contributing to the theory of AHP in other aspects. The developed framework starts with the \textit{true weight vector} in order to construct the numerical pairwise comparison matrices which are used to determine the priority vectors. As part of this framework, two new performance metrics are introduced to the literature (i.e., GCIV-AV and GCIV-VW) alongside with the commonly used CIV (i.e., GCIV-AW). GCIV-AV addresses the compatibility with the numerical pairwise comparison matrix and the theoretical numerical pairwise comparison matrix that is constructed from the true priority vectors. On the other hand, GCIV-VW addresses the compatibility between the  true priority vector and the calculated priority vector.

The second contribution of this study is the BAGINS that can be used to generate individualized scales. Unlike the existing algorithms that are proposed earlier for this purpose, BAGINS doesn't target the inconsistency of the numerical comparison matrix but aims to choose the individualized scales that would improve the compatibility of the numerical comparison matrix and the theoretical numerical comparison matrix. 

In order to measure the performance of BAGINS, a numerical study and two experimental analyses were conducted. Both the numerical and empirical data sets, with necessary Matlab codes and documentation are made available online for the use of researchers and practitioners. We sincerely believe the benefit of open data and consider it as the third contribution of this research. 

The numerical study demonstrated that methods based on the individualized scales outperform the conventional Saaty method with fixed scale statistically significantly in 11 out of the 14 experimental conditions in terms of the newly introduced GCIV-AV metric and the conventional GCIV-AW metrics (and for the remaining three experimental conditions the individualized scales outperforms the fixed sale \textit{on the average}). On the other hand, among the two methods that are based on individualized scales, NLP statistically significantly outperforms BAGINS in terms of the GCIV-AW metric, and BAGINS statistically significantly outperforms NLP in terms of the GCIV-VW metric for all of the seven experimental conditions, and in terms of the GCIV-AV, this was the case for five out of the seven experimental conditions.

One of the interesting observations from the numerical study which might worth further analysis is the performance of Saaty with fixed scale in terms of the GCIV-VW metric. As depicted earlier in Figure \ref{GCIVdiagram}, in terms of the compatibility between the theoretical true weights and the numerical pairwise comparison matrix, the fixed scale based Saaty is not performing as good as the individualized scaled based approaches. This is also the case for the compatibility between the numerical pairwise comparison matrix and the theoretical numerical matrix constructed from the calculated priority vector. However, when one considers the process end-to-end, Saaty outperforms the other two methods. That is to say, it somehow straightens out the bad performances in the intermediary two steps (i.e., Step 1: Representation of the true weights with the numerical pairwise comparison matrix, and Step 2: Determining the priority vector from the numerical pairwise comparison matrix). We left this observation as a possible research question.

Some other future research questions are regarding the improvement of the BAGINS. As is, it targets a proxy optimization problem that minimizes a squared deviation as explained earlier. Another option might be targeting the absolute deviation. A more interesting question would be changing the method that is used to determine the priority vector. As is, BAGINs use Saaty's conventional eigenvalue-eigenvector approach. However one might also consider other methods (e.g., LLSM, Mean of Normalized Values Heuristic, etc.). It is also possible to consider the recursive version of BAGINS in which after the determination of the individualized scale, the priority vector can be recalculated and the whole process is repeated until a termination criterion. The convergence of the recursive BAGINS is also another possible research problem that might be addressed in the future. Finally, additional \textit{empirical} studies would benefit not only this research but also other research that contributes to the theory of AHP (e.g., new algorithms to determine priority vectors, new performance metrics for evaluation of different approaches, new static and/or individualized scales, etc.).

\bibliographystyle{model5-names}

\section*{References}
\bibliography{BAGINS}

\begin{thebibliography}{36}
\expandafter\ifx\csname natexlab\endcsname\relax\def\natexlab#1{#1}\fi
\providecommand{\url}[1]{\texttt{#1}}
\providecommand{\href}[2]{#2}
\providecommand{\path}[1]{#1}
\providecommand{\DOIprefix}{doi:}
\providecommand{\ArXivprefix}{arXiv:}
\providecommand{\URLprefix}{URL: }
\providecommand{\Pubmedprefix}{pmid:}
\providecommand{\doi}[1]{\href{http://dx.doi.org/#1}{\path{#1}}}
\providecommand{\Pubmed}[1]{\href{pmid:#1}{\path{#1}}}
\providecommand{\bibinfo}[2]{#2}
\ifx\xfnm\relax \def\xfnm[#1]{\unskip,\space#1}\fi
\bibitem[{Ahmed(2022)}]{ahmedfaranGitHub}
\bibinfo{author}{Ahmed, F.} (\bibinfo{year}{2022}).
\newblock \bibinfo{title}{{Individualized-Numerical-Scales}}.
\newblock \URLprefix
  \url{https://github.com/ahmedfaran/Individualized-Numerical-Scales}
  \bibinfo{note}{[Online; accessed 13. Sep. 2020]}.
\bibitem[{Brunelli(2018)}]{brunelli2018survey}
\bibinfo{author}{Brunelli, M.} (\bibinfo{year}{2018}).
\newblock \bibinfo{title}{A survey of inconsistency indices for pairwise
  comparisons}.
\newblock {\it \bibinfo{journal}{International Journal of General Systems}\/},
  {\it \bibinfo{volume}{47}\/}, \bibinfo{pages}{751--771}.
\bibitem[{Budescu \& Wallsten(1985)}]{budescu1985consistency}
\bibinfo{author}{Budescu, D.~V.}, \& \bibinfo{author}{Wallsten, T.~S.}
  (\bibinfo{year}{1985}).
\newblock \bibinfo{title}{Consistency in interpretation of probabilistic
  phrases}.
\newblock {\it \bibinfo{journal}{Organizational Behavior and Human Decision
  Processes}\/},  {\it \bibinfo{volume}{36}\/}, \bibinfo{pages}{391--405}.
\bibitem[{Cavallo et~al.(2019)Cavallo, Ishizaka, Olivieri \&
  Squillante}]{cavallo2019comparing}
\bibinfo{author}{Cavallo, B.}, \bibinfo{author}{Ishizaka, A.},
  \bibinfo{author}{Olivieri, M.~G.}, \& \bibinfo{author}{Squillante, M.}
  (\bibinfo{year}{2019}).
\newblock \bibinfo{title}{Comparing inconsistency of pairwise comparison
  matrices depending on entries}.
\newblock {\it \bibinfo{journal}{Journal of the Operational Research
  Society}\/},  {\it \bibinfo{volume}{70}\/}, \bibinfo{pages}{842--850}.
\bibitem[{Choo et~al.(2016)Choo, Wedley \& Wijnmalen}]{choo2016mathematical}
\bibinfo{author}{Choo, E.~U.}, \bibinfo{author}{Wedley, W.~C.}, \&
  \bibinfo{author}{Wijnmalen, D.~J.} (\bibinfo{year}{2016}).
\newblock \bibinfo{title}{Mathematical support for the geometric mean when
  deriving a consistent matrix from a pairwise ratio matrix}.
\newblock {\it \bibinfo{journal}{Fundamenta Informaticae}\/},  {\it
  \bibinfo{volume}{144}\/}, \bibinfo{pages}{263--278}.
\bibitem[{Crawford \& Williams(1985)}]{crawford1985note}
\bibinfo{author}{Crawford, G.}, \& \bibinfo{author}{Williams, C.}
  (\bibinfo{year}{1985}).
\newblock \bibinfo{title}{A note on the analysis of subjective judgment
  matrices}.
\newblock {\it \bibinfo{journal}{Journal of Mathematical Psychology}\/},  {\it
  \bibinfo{volume}{29}\/}, \bibinfo{pages}{387--405}.
\bibitem[{Dodd \& Donegan(1995)}]{dodd1995comparison}
\bibinfo{author}{Dodd, F.}, \& \bibinfo{author}{Donegan, H.}
  (\bibinfo{year}{1995}).
\newblock \bibinfo{title}{Comparison of prioritization techniques using
  interhierarchy mappings}.
\newblock {\it \bibinfo{journal}{Journal of the Operational Research
  Society}\/},  {\it \bibinfo{volume}{46}\/}, \bibinfo{pages}{492--498}.
\bibitem[{Dong et~al.(2013)Dong, Hong, Xu \& Yu}]{dong2013numerical}
\bibinfo{author}{Dong, Y.}, \bibinfo{author}{Hong, W.-C.}, \bibinfo{author}{Xu,
  Y.}, \& \bibinfo{author}{Yu, S.} (\bibinfo{year}{2013}).
\newblock \bibinfo{title}{Numerical scales generated individually for analytic
  hierarchy process}.
\newblock {\it \bibinfo{journal}{European Journal of Operational Research}\/},
  {\it \bibinfo{volume}{229}\/}, \bibinfo{pages}{654--662}.
\bibitem[{Emrouznejad \& Marra(2017)}]{emrouznejad2017state}
\bibinfo{author}{Emrouznejad, A.}, \& \bibinfo{author}{Marra, M.}
  (\bibinfo{year}{2017}).
\newblock \bibinfo{title}{The state of the art development of ahp (1979--2017):
  A literature review with a social network analysis}.
\newblock {\it \bibinfo{journal}{International journal of production
  research}\/},  {\it \bibinfo{volume}{55}\/}, \bibinfo{pages}{6653--6675}.
\bibitem[{Finan \& Hurley(1999)}]{finan1999transitive}
\bibinfo{author}{Finan, J.~S.}, \& \bibinfo{author}{Hurley, W.~J.}
  (\bibinfo{year}{1999}).
\newblock \bibinfo{title}{Transitive calibration of the ahp verbal scale}.
\newblock {\it \bibinfo{journal}{European Journal of Operational Research}\/},
  {\it \bibinfo{volume}{112}\/}, \bibinfo{pages}{367--372}.
\bibitem[{Hacioglu et~al.(2021)Hacioglu, Chlyeh, Yilmaz, Tatoglu \&
  Delen}]{hacioglu2021crafting}
\bibinfo{author}{Hacioglu, U.}, \bibinfo{author}{Chlyeh, D.},
  \bibinfo{author}{Yilmaz, M.~K.}, \bibinfo{author}{Tatoglu, E.}, \&
  \bibinfo{author}{Delen, D.} (\bibinfo{year}{2021}).
\newblock \bibinfo{title}{Crafting performance-based cryptocurrency mining
  strategies using a hybrid analytics approach}.
\newblock {\it \bibinfo{journal}{Decision Support Systems}\/},  {\it
  \bibinfo{volume}{142}\/}, \bibinfo{pages}{113473}.
\bibitem[{Hajiali et~al.(2022)Hajiali, Teimoury, Rabiee \&
  Delen}]{hajiali2022interactive}
\bibinfo{author}{Hajiali, M.}, \bibinfo{author}{Teimoury, E.},
  \bibinfo{author}{Rabiee, M.}, \& \bibinfo{author}{Delen, D.}
  (\bibinfo{year}{2022}).
\newblock \bibinfo{title}{An interactive decision support system for real-time
  ambulance relocation with priority guidelines}.
\newblock {\it \bibinfo{journal}{Decision Support Systems}\/},  {\it
  \bibinfo{volume}{155}\/}, \bibinfo{pages}{113712}.
\bibitem[{Harker \& Vargas(1987)}]{harker1987theory}
\bibinfo{author}{Harker, P.~T.}, \& \bibinfo{author}{Vargas, L.~G.}
  (\bibinfo{year}{1987}).
\newblock \bibinfo{title}{The theory of ratio scale estimation: Saaty's
  analytic hierarchy process}.
\newblock {\it \bibinfo{journal}{Management Science}\/},  {\it
  \bibinfo{volume}{33}\/}, \bibinfo{pages}{1383--1403}.
\bibitem[{Herrera \& Mart{\'\i}nez(2000)}]{herrera20002}
\bibinfo{author}{Herrera, F.}, \& \bibinfo{author}{Mart{\'\i}nez, L.}
  (\bibinfo{year}{2000}).
\newblock \bibinfo{title}{A 2-tuple fuzzy linguistic representation model for
  computing with words}.
\newblock {\it \bibinfo{journal}{IEEE Transactions on fuzzy systems}\/},  {\it
  \bibinfo{volume}{8}\/}, \bibinfo{pages}{746--752}.
\bibitem[{Huizingh \& Vrolijk(1997)}]{huizingh1997comparison}
\bibinfo{author}{Huizingh, E.~K.}, \& \bibinfo{author}{Vrolijk, H.~C.}
  (\bibinfo{year}{1997}).
\newblock \bibinfo{title}{A comparison of verbal and numerical judgments in the
  analytic hierarchy process}.
\newblock {\it \bibinfo{journal}{Organizational Behavior and Human Decision
  Processes}\/},  {\it \bibinfo{volume}{70}\/}, \bibinfo{pages}{237--247}.
\bibitem[{Ishizaka et~al.(2011)Ishizaka, Balkenborg \&
  Kaplan}]{ishizaka2011influence}
\bibinfo{author}{Ishizaka, A.}, \bibinfo{author}{Balkenborg, D.}, \&
  \bibinfo{author}{Kaplan, T.} (\bibinfo{year}{2011}).
\newblock \bibinfo{title}{Influence of aggregation and measurement scale on
  ranking a compromise alternative in ahp}.
\newblock {\it \bibinfo{journal}{Journal of the Operational Research
  Society}\/},  {\it \bibinfo{volume}{62}\/}, \bibinfo{pages}{700--710}.
\bibitem[{Ishizaka \& Labib(2011)}]{ishizaka2011review}
\bibinfo{author}{Ishizaka, A.}, \& \bibinfo{author}{Labib, A.}
  (\bibinfo{year}{2011}).
\newblock \bibinfo{title}{Review of the main developments in the analytic
  hierarchy process}.
\newblock {\it \bibinfo{journal}{Expert Systems with Applications}\/},  {\it
  \bibinfo{volume}{38}\/}, \bibinfo{pages}{14336--14345}.
\bibitem[{Kahneman \& Tversky(1979)}]{Kahneman1979Mar}
\bibinfo{author}{Kahneman, D.}, \& \bibinfo{author}{Tversky, A.}
  (\bibinfo{year}{1979}).
\newblock \bibinfo{title}{{Prospect Theory: An Analysis of Decision under Risk
  on JSTOR}}.
\newblock {\it \bibinfo{journal}{Econometrica}\/},  {\it
  \bibinfo{volume}{47}\/}, \bibinfo{pages}{263--292}. \URLprefix
  \url{https://www.jstor.org/stable/1914185}.
\bibitem[{Kent(1964)}]{kent1964words}
\bibinfo{author}{Kent, S.} (\bibinfo{year}{1964}).
\newblock \bibinfo{title}{Words of estimative probability}.
\newblock {\it \bibinfo{journal}{Studies in intelligence}\/},  {\it
  \bibinfo{volume}{8}\/}, \bibinfo{pages}{49--65}.
\bibitem[{Li et~al.(2021)Li, Hsieh, Lin \& Wei}]{li2021social}
\bibinfo{author}{Li, Y.-M.}, \bibinfo{author}{Hsieh, C.-Y.},
  \bibinfo{author}{Lin, L.-F.}, \& \bibinfo{author}{Wei, C.-H.}
  (\bibinfo{year}{2021}).
\newblock \bibinfo{title}{A social mechanism for task-oriented crowdsourcing
  recommendations}.
\newblock {\it \bibinfo{journal}{Decision Support Systems}\/},  {\it
  \bibinfo{volume}{141}\/}, \bibinfo{pages}{113449}.
\bibitem[{Liang et~al.(2008)Liang, Wang, Hua \& Zhang}]{liang2008mapping}
\bibinfo{author}{Liang, L.}, \bibinfo{author}{Wang, G.}, \bibinfo{author}{Hua,
  Z.}, \& \bibinfo{author}{Zhang, B.} (\bibinfo{year}{2008}).
\newblock \bibinfo{title}{Mapping verbal responses to numerical scales in the
  analytic hierarchy process}.
\newblock {\it \bibinfo{journal}{Socio-Economic Planning Sciences}\/},  {\it
  \bibinfo{volume}{42}\/}, \bibinfo{pages}{46--55}.
\bibitem[{Lootsma(1993)}]{lootsma1993scale}
\bibinfo{author}{Lootsma, F.~A.} (\bibinfo{year}{1993}).
\newblock \bibinfo{title}{Scale sensitivity in the multiplicative ahp and
  smart}.
\newblock {\it \bibinfo{journal}{Journal of Multi-Criteria Decision
  Analysis}\/},  {\it \bibinfo{volume}{2}\/}, \bibinfo{pages}{87--110}.
\bibitem[{Mandel et~al.(2021)Mandel, Dhami, Tran \&
  Irwin}]{mandel2021arithmetic}
\bibinfo{author}{Mandel, D.~R.}, \bibinfo{author}{Dhami, M.~K.},
  \bibinfo{author}{Tran, S.}, \& \bibinfo{author}{Irwin, D.}
  (\bibinfo{year}{2021}).
\newblock \bibinfo{title}{Arithmetic computation with probability words and
  numbers}.
\newblock {\it \bibinfo{journal}{Journal of Behavioral Decision Making}\/}, .
\bibitem[{Morgenstern \& Von~Neumann(1953)}]{morgenstern1953theory}
\bibinfo{author}{Morgenstern, O.}, \& \bibinfo{author}{Von~Neumann, J.}
  (\bibinfo{year}{1953}).
\newblock {\it \bibinfo{title}{Theory of Games and Economic Behavior}\/}.
\newblock \bibinfo{publisher}{Princeton university press}.
\bibitem[{P{\"o}yh{\"o}nen et~al.(1997)P{\"o}yh{\"o}nen, HAeMAeLAeINEN \&
  Salo}]{poyhonen1997experiment}
\bibinfo{author}{P{\"o}yh{\"o}nen, M.~A.}, \bibinfo{author}{HAeMAeLAeINEN,
  R.~P.}, \& \bibinfo{author}{Salo, A.~A.} (\bibinfo{year}{1997}).
\newblock \bibinfo{title}{An experiment on the numerical modelling of verbal
  ratio statements}.
\newblock {\it \bibinfo{journal}{Journal of Multi-Criteria Decision
  Analysis}\/},  {\it \bibinfo{volume}{6}\/}, \bibinfo{pages}{1--10}.
\bibitem[{Saaty(1994)}]{saaty1994ratio}
\bibinfo{author}{Saaty, T.} (\bibinfo{year}{1994}).
\newblock \bibinfo{title}{A ratio scale metric and the compatibility of ratio
  scales: The possibility of arrow's impossibility theorem}.
\newblock {\it \bibinfo{journal}{Applied Mathematics Letters}\/},  {\it
  \bibinfo{volume}{7}\/}, \bibinfo{pages}{51--57}.
\bibitem[{Saaty(1977)}]{saaty1977scaling}
\bibinfo{author}{Saaty, T.~L.} (\bibinfo{year}{1977}).
\newblock \bibinfo{title}{A scaling method for priorities in hierarchical
  structures}.
\newblock {\it \bibinfo{journal}{Journal of Mathematical Psychology}\/},  {\it
  \bibinfo{volume}{15}\/}, \bibinfo{pages}{234--281}.
\bibitem[{Saaty \& Tran(2007)}]{saaty2007invalidity}
\bibinfo{author}{Saaty, T.~L.}, \& \bibinfo{author}{Tran, L.~T.}
  (\bibinfo{year}{2007}).
\newblock \bibinfo{title}{On the invalidity of fuzzifying numerical judgments
  in the analytic hierarchy process}.
\newblock {\it \bibinfo{journal}{Mathematical and Computer Modelling}\/},  {\it
  \bibinfo{volume}{46}\/}, \bibinfo{pages}{962--975}.
\bibitem[{Salo \& H{\"a}m{\"a}l{\"a}inen(1997)}]{salo1997measurement}
\bibinfo{author}{Salo, A.~A.}, \& \bibinfo{author}{H{\"a}m{\"a}l{\"a}inen,
  R.~P.} (\bibinfo{year}{1997}).
\newblock \bibinfo{title}{On the measurement of preferences in the analytic
  hierarchy process}.
\newblock {\it \bibinfo{journal}{Journal of Multi-Criteria Decision
  Analysis}\/},  {\it \bibinfo{volume}{6}\/}, \bibinfo{pages}{309--319}.
\bibitem[{Sato \& Tan(2022)}]{sato2022inconsistency}
\bibinfo{author}{Sato, Y.}, \& \bibinfo{author}{Tan, K.~H.}
  (\bibinfo{year}{2022}).
\newblock \bibinfo{title}{Inconsistency indices in pairwise comparisons: an
  improvement of the consistency index}.
\newblock {\it \bibinfo{journal}{Annals of Operations Research}\/},  (pp.
  \bibinfo{pages}{1--22}).
\bibitem[{Tanoumand et~al.(2017)Tanoumand, Ozdemir, Kilic \&
  Ahmed}]{tanoumand2017selecting}
\bibinfo{author}{Tanoumand, N.}, \bibinfo{author}{Ozdemir, D.~Y.},
  \bibinfo{author}{Kilic, K.}, \& \bibinfo{author}{Ahmed, F.}
  (\bibinfo{year}{2017}).
\newblock \bibinfo{title}{Selecting cloud computing service provider with fuzzy
  ahp}.
\newblock In {\it \bibinfo{booktitle}{2017 IEEE international conference on
  fuzzy systems (FUZZ-IEEE)}\/} (pp. \bibinfo{pages}{1--5}).
\newblock \bibinfo{organization}{IEEE}.
\bibitem[{Tavana et~al.(1997)Tavana, Kennedy \& Mohebbi}]{tavana1997applied}
\bibinfo{author}{Tavana, M.}, \bibinfo{author}{Kennedy, D.~T.}, \&
  \bibinfo{author}{Mohebbi, B.} (\bibinfo{year}{1997}).
\newblock \bibinfo{title}{An applied study using the analytic hierarchy process
  to translate common verbal phrases to numerical probabilities}.
\newblock {\it \bibinfo{journal}{Journal of Behavioral Decision Making}\/},
  {\it \bibinfo{volume}{10}\/}, \bibinfo{pages}{133--150}.
\bibitem[{Veisi et~al.(2022)Veisi, Deihimfard, Shahmohammadi \&
  Hydarzadeh}]{veisi2022application}
\bibinfo{author}{Veisi, H.}, \bibinfo{author}{Deihimfard, R.},
  \bibinfo{author}{Shahmohammadi, A.}, \& \bibinfo{author}{Hydarzadeh, Y.}
  (\bibinfo{year}{2022}).
\newblock \bibinfo{title}{Application of the analytic hierarchy process (ahp)
  in a multi-criteria selection of agricultural irrigation systems}.
\newblock {\it \bibinfo{journal}{Agricultural Water Management}\/},  {\it
  \bibinfo{volume}{267}\/}, \bibinfo{pages}{107619}.
\bibitem[{Vlaev et~al.(2011)Vlaev, Chater, Stewart \& Brown}]{vlaev2011does}
\bibinfo{author}{Vlaev, I.}, \bibinfo{author}{Chater, N.},
  \bibinfo{author}{Stewart, N.}, \& \bibinfo{author}{Brown, G.~D.}
  (\bibinfo{year}{2011}).
\newblock \bibinfo{title}{Does the brain calculate value?}
\newblock {\it \bibinfo{journal}{Trends in cognitive sciences}\/},  {\it
  \bibinfo{volume}{15}\/}, \bibinfo{pages}{546--554}.
\bibitem[{WEP(2020)}]{WEPmedecine}
\bibinfo{author}{WEP} (\bibinfo{year}{2020}).
\newblock \bibinfo{title}{{The Numbers in Our Words: Words of Estimative
  Probability {$\vert$} Lone Gunman}}.
\newblock \URLprefix
  \url{https://www.lonegunman.co.uk/2011/02/03/the-numbers-in-our-words-words-of-estimative-probability}
  \bibinfo{note}{[Online; accessed 15. Sep. 2020]}.
\bibitem[{Zhou et~al.(2018)Zhou, Dong, Zhang \& Gao}]{zhou2018analytic}
\bibinfo{author}{Zhou, Q.}, \bibinfo{author}{Dong, Y.}, \bibinfo{author}{Zhang,
  H.}, \& \bibinfo{author}{Gao, Y.} (\bibinfo{year}{2018}).
\newblock \bibinfo{title}{The analytic hierarchy process with personalized
  individual semantics}.
\newblock {\it \bibinfo{journal}{International Journal of Computational
  Intelligence Systems}\/},  {\it \bibinfo{volume}{11}\/},
  \bibinfo{pages}{451--468}.

\end{thebibliography}

\end{document}